\documentstyle[12pt] {article}
\def\zid{1\kern-0.36em\llap~1}

\newcommand{\beq}{\begin{equation}}
\newcommand{\ber}{\begin{eqnarray}}
\newcommand{\eeq}{\end{equation}}
\newcommand{\eer}{\end{eqnarray}}

\begin{document}

\begin{titlepage}
\rightline{SUNY BING 5/28/97R2, hep-ph/9707211}
\vspace{1mm}
\begin{center}
{\bf GENERAL TESTS FOR $t \rightarrow W^+ b $ COUPLINGS AT
HADRON COLLIDERS}\\
\vspace{2mm}
Charles A. Nelson\footnote{Electronic address: cnelson @
bingvmb.cc.binghamton.edu }, Brian T. Kress, Marco Lopes, and
Thomas P. McCauley\\
{\it Department of Physics, State University of New York at
Binghamton\\
Binghamton, N.Y. 13902-6016}\\[2mm]
\end{center}


\begin{abstract}

The modularity property of the helicity formalism is used to
provide amplitude expressions and stage-two spin-correlation
functions which can easily be used in direct experimental
searches for electro-weak symmetries and dynamics in the
decay processes $t \rightarrow W^+ b$, $\bar t \rightarrow
W^- \bar b$.  The formalism is used to describe the decay
sequences
$t\rightarrow
W^{+}b\rightarrow (l^{+}\nu )b$, and $t\rightarrow
W^{+}b\rightarrow (j_{\bar
d}j_u)b$. Helicity amplitudes for $t \rightarrow W^+ b $\ are
obtained for the
most general $J_{\bar b t}$ current.  Thereby, the most
general
Lorentz-invariant decay-density-matrix
for $t\rightarrow
W^{+}b\rightarrow (l^{+}\nu )b$, or for $t\rightarrow
W^{+}b\rightarrow (j_{\bar
d}j_u)b$, is expressed in terms of eight helicity parameters
and,
equivalently, in terms of the structures of the $J_{\bar b
t}$
current.
The parameters
are physically defined in terms of
partial-width-intensities for
polarized-final-states in \newline $t\rightarrow
W^{+}b$ decay. The full angular distribution for the
reactions  $q \bar
q$ and $g g \rightarrow t \bar t \rightarrow (W^+ b) (W^-
\bar b)
\rightarrow \ldots$ can be used to measure these parameters.
Since
this adds on spin-correlation information from the next stage
of
decays in the decay sequence, such an energy-angular
distribution is
called a stage-two spin-correlation (S2SC) function.

\end{abstract}

\end{titlepage}

\section{INTRODUCTION}

While in the standard model the violations of $CP, T$, and
$(V-A)$ symmetry are phenomenologically well-described by the
Higgs mechanism and the CKM matrix, the depth of the
dynamical
understanding remains open to question.  In particular, the
Yukawa couplings of the fermions, and the CKM mixing angles
and CP phase parameter are inserted by-hand. For this reason,
and the new fermionic mass scale of $\sim 175$ $GeV$ provided
by the
recently discovered top quark[1-3],  it is important to probe
for
new and/or additional symmetry violations at $m_t
\sim 175$ $GeV$.

We use the modularity property of the helicity
formalism\cite{JW} to
provide amplitude expressions and stage-two spin-correlation
functions which can easily be used in direct experimental
searches for electro-weak symmetries and dynamics in the
decay processes $t \rightarrow W^+ b$, $\bar t \rightarrow
W^- \bar b$.
Stage-two spin-correlation functions are also a
useful technique for testing the the symmetry properties and
dynamics of $t \bar t$ pair
production in both the  $q \bar q \rightarrow t \bar t $
channel and the $g g \rightarrow t \bar t $
channel[5,6].

The
reader should be aware that it is not necessary to use
the helicity formalism \cite{JW} because the observables are
physically defined in terms $t \rightarrow W^+ b$ decay
partial width
intensities for polarized-final-states. However, the helicity
formalism does provide a lucid, flexible, physical
framework for connecting
Lorentz-invariant couplings at the Lagrangian level with
Lorentz-invariant spin-correlation functions.
In practice, the helicity formalism also frequently provides
insights and easy checks on the
resulting formulas.

The literature on polarimetry methods and spin-correlation
functions in $t$ quark physics includes Refs.[5,7,6].
Literature on methods to test for $CP$
violation in $t$
reactions includes Refs.[5,8,9,6].

In this paper, we concentrate on the most general
Lorentz-invariant decay-density-matrix
$R_{ \lambda_1,\lambda_1^{\prime} }$
for $t \rightarrow
W^{+}b\rightarrow (l^{+}\nu )b$, or for $t \rightarrow
W^{+}b\rightarrow (j_{\bar
d}j_u)b$ where $\lambda_1, \lambda_1^{\prime}$ $ = \pm 1/2$
is the
$t$
helicity. \newline
$R_{ \lambda_1,\lambda_1^{\prime} }$ is expressed in terms of
eight
helicity
parameters[6,10].
The diagonal elements are simply the angular
distributions $
\frac{dN}{d(\cos
\theta
_1^t )d(\cos \tilde \theta _a)d \tilde \phi_a} $ for the
polarized $t$ decay
chain, $t\rightarrow
W^{+}b\rightarrow (l^{+}\nu )b$, or $t\rightarrow
W^{+}b\rightarrow (j_{\bar
d}j_u)b$.

There are eight $t \rightarrow W^+ b$ decay parameters since
there are the four $W_{L,T}$ $b_{L,R}$ final-state
combinations:
The first parameter is simply $\Gamma \equiv
\Gamma
_L^{+}+\Gamma _T^{+}$, i.e. the partial width
for $t\rightarrow W^{+} b$.
The subscripts on
the $\Gamma $'s denote the
polarization of the final $W^{+}$, either
``L=longitudinal'' or
``T=transverse''; superscripts denote ``$\pm $ for
sum/difference of
the $b_{L\ }$versus $b_R$ contributions''.  In terms of the
helicity amplitudes defined in Sec. 2,
\begin{equation}
\begin{array}{c}
\Gamma _L^{\pm }=\left| A(0,-\frac 12)\right| ^2\pm \left|
A(0,\frac
12)\right| ^2 \\
\Gamma _T^{\pm }=\left| A(-1,-\frac 12)\right| ^2\pm \left|
A(1,\frac
12)\right| ^2.
\end{array}
\end{equation}
Such final-state-polarized
partial widths are observables and, indeed, the
equivalent helicity parameters $\xi , \sigma , \ldots $ can
be
measured by various polarimetry and
spin-correlation
techniques.

The second helicity parameter is the $b$ quark's
chirality
parameter $\xi \equiv \frac 1\Gamma (\Gamma _L^{-}+\Gamma
_T^{-})$.
Equivalently, \newline  \hspace{1pc} $  \xi \equiv$ (Prob
$b$ is
$b_L$) $ - $
(Prob $b$ is $b_R$),
\begin{equation}
\xi \equiv |< b_L | b >|^{2} - |< b_R
| b
>|^{2}
\end{equation}
So for $m_b=0$, a value $\xi = 1$ means the coupled $b$ quark
is
pure $b_L$, i.e. $\lambda_b = -1/2$.  For $m_b=4.5 GeV$,
$\xi=0.9993$ for a pure $V-A$ coupling\cite{PL}.

The remaining two
partial-width parameters are defined by
\begin{equation}
\zeta \equiv (\Gamma _L^{-}-\Gamma _T^{-})/ \Gamma ,
\hspace{1pc} \sigma \equiv (\Gamma
_L^{+}-\Gamma
_T^{+})/ \Gamma .
\end{equation}
This implies for $W^+$ polarimetry that  \newline  $
\hspace{2pc}
\sigma
=$ (Prob
$W^{+}$ is $W_L$)
$
-
$ (Prob $W^{+}$ is $W_T$), \newline is the analogue of
the
$b$ quark's chirality parameter in
Eq.(2).
Thus, the parameter $\sigma$ measures
the
degree of polarization, ``L minus T", of the emitted $W^+$.
For a pure $(V - A)$, or $(V + A)$, coupling and the
empirical masses, $\sigma = 0.4057$.
The ``pre-SSB" parameter $\zeta =0.4063 $ characterizes the
remaining
odd-odd
mixture of the $b$ and $W^+$ spin-polarizations.

To describe the interference between the $W_L$ and
$W_R$ amplitudes, we define the
four normalized parameters,
\begin{equation}
\begin{array}{c}
\omega \equiv I_{
{\cal R}}^{-}\ / \Gamma , \hspace{2pc}  \eta
\equiv I_{
{\cal R}}^{+}\ / \Gamma  \\
\omega ^{\prime
}\equiv I_{
{\cal I}}^{-}\ / \Gamma , \hspace{2pc} \eta
^{\prime }\equiv
I_{{\cal I}%
}^{+}\ / \Gamma
\end{array}
\end{equation}
The associated $W_L - W_T$ interference intensities are
\begin{equation}
\begin{array}{c}
I_{{\cal R}}^{\pm }=\left| A(0,-\frac 12)\right| \left| A(-
1,-\frac
12)\right| \cos \beta _a  \cr \pm \left| A(0,\frac
12)\right|
\left| A(1,\frac
12)\right| \cos \beta _a^R  \\
I_{{\cal I}}^{\pm }=\left| A(0,-\frac 12)\right| \left| A(-
1,-\frac
12)\right| \sin \beta _a \cr \pm \left| A(0,\frac
12)\right|
\left| A(1,\frac
12)\right| \sin \beta _a^R
\end{array}
\end{equation}
Here,  $\beta _a\equiv
\phi _{-1}^a-\phi _0^a$, and $\beta
_a^R\equiv \phi
_1^a-\phi _0^{aR}$\ are the measurable phase differences of
of the
associated helicity amplitudes
$A(\lambda_{W^+},\lambda_b)=\left|
A\right| \exp \iota \phi $ in the standard helicity amplitude
phase convention\cite{JW}. For the empirical masses,
$\omega=0.4566$ and $ \eta=0.4568$ are also unequal since
$m_b= 4.5 GeV$.  If unlike in the SM $\beta_a^R \neq 0$, then
from Eq.(5) there are the inequalities $\omega^{\prime} \neq
\eta^{\prime}$ and $\omega \neq \eta$, but both of these
inequalities will be insignificant versus anticipated
empirical precisions unless both $b_R$ amplitudes, $\lambda_W
= 0,1$, are unexpectedly enhanced.

If one factors out ``W-polarimetry factors", see below, via $
\sigma = {\cal S}_W \tilde
\sigma$,
$\omega = {\cal R}_W \tilde \omega$, $\ldots$ the
parameters all equal one or zero for a pure $(V-A)$ coupling
and $m_b =0$ ( $\omega^{\prime
} $ $= \eta^{\prime
}$ $= 0$ ).

{\bf Important Remarks:}

(1) The analytic forms of ``$\xi, \sigma, \zeta,  \ldots $"
are very
distinct for different unique Lorentz couplings, see Table 1.
This is also true for the partial-width-intensities for
polarized-final-states, see Table 2. This is
indicative of the analyzing power of stage-two
spin-correlation techniques
for analyzing $t \rightarrow W^+ b$ decay. Both the real and
the imaginary parts of the associated helicity amplitudes can
be directly measured.

(2) Primed parameters $ \omega ^{\prime } \neq 0 $ and/or $
\eta
^{\prime }
\neq 0
\Longrightarrow  \tilde{T}_{FS} $ is violated. $\tilde
T_{FS}$
invariance will be violated when either there is a violation
of
canonical $T$ invariance or when there are absorptive
final-state
interactions.

(3) Barred parameters $ \bar{\xi}, \bar{\zeta}, \ldots $ have
the analogous
definitions for the $CP$ conjugate process, $\bar t
\rightarrow W^- \bar b
$. Therefore, any $
\bar{\xi} \neq \xi,
\bar{\zeta} \neq \zeta, \ldots $ $ \Longrightarrow $ CP is
violated.  That is, ``slashed parameters" $\not\xi \equiv \xi
-
\bar{\xi}$, \ldots, could be introduced to characterize and
quantify the degree of CP violation.  This should be regarded
as a test for the presence of a non-CKM-type CP violation
because, normally, a  CKM-phase
will contribute equally at tree level to both the $t
\rightarrow W^+ b_L$
decay amplitudes and so a CKM-phase will cancel out in the
ratio of their
moduli and in their relative phase.  There are four tests for
non-CKM-type CP violation[6,11].

(4). These helicity parameters appear in the general angular
distributions for
the polarized $t\rightarrow
W^{+}b\rightarrow (l^{+}\nu )b$ decay chain, and for
$t\rightarrow
W^{+}b\rightarrow (j_{\bar
d}j_u)b$. Such formulas for the associated ``stage-two
spin-correlation"
(S2SC) functions in terms of
these eight helicity parameters are derived below in Sec. 5.

(5) In the presence of additional Lorentz structures, ``W-
polarimetry factors" ${\cal S}_W= 0.4068$ and ${\cal R}_W=
0.4567$ naturally appear[5,10] because of the referencing of
``new physics" to the $(V-A)$ structure of the SM. These
important factors are
\beq
{\cal S}_W =\frac{1-2\frac{m_W
^2}{m_t ^2}}{1+2\frac{m_W ^2}{m_t ^2}}
\eeq
and
\beq
{\cal R}_W =\frac{\sqrt{2}\frac{m_W}{m_t}}
{1+2\frac{m_W^2}{m_t ^2}}
\eeq

We have introduced ${\cal S}_W $ and
${\cal R}_W
$ because we are analyzing versus a
reference ${J_{\bar b t} }$
theory consisting of ``a
mixture of only $V$ and $A$ couplings with $m_{b} = 0$".
For the third generation of quarks and leptons, this is the
situation in the SM before the Higgs mechanism is invoked.
We refer to this limit as the ``pre-SSB" case.  In this case,
these W-polarimetry factors have a simple physical
interpretation:
for  $t \rightarrow W_{L,T} ^+ b $ the factor
\newline
$ {\cal S}_W = $ (Prob $W_L $)
$ - $
(Prob $W_T $), and the factor
\newline
$ {\cal R}_W = $ the ``geometric
mean of these
probablities"
$= \sqrt{(Prob \hspace{.1pc} W_L ) (Prob
\hspace{.1pc}
W_T )}.$
\newline
These factors are not independent since
$ ({\cal S}_W)^2 + 4 ({\cal R}_W)^2 = 1 $.
[ If experiments for the lighter quarks and leptons had
suggested instead a different dominant
Lorentz-structure than $V-A$, say \newline ``$ f_M+f_E$ '',
then per
Table 1 we would have replaced ${\cal S}_W $ everywhere by
$
(-2+\frac{w^2}{t^2})/(2+\frac{w^2}{t^2})$,
etc. ].

In the ``pre-SSB" case, each of the eight helicity parameters
also has a simple probabilistic significance for they
are
each directly proportional to $\Gamma ,\xi ,
{\cal S}_W $, or ${\cal R}_W $: $\sigma \rightarrow
{\cal S}_W,  \zeta
\rightarrow {\cal S}_W \xi , \omega \rightarrow {\cal R}_W
\xi, \eta \rightarrow {\cal R}_W  $. {\bf Therefore},
precision measurements with $\xi$ and $\zeta$ distinct, and
with $\xi$ and $\omega$ distinct, {\bf will be two useful
probes} of the dynamics of EW spontaneous symmetry breaking,
see Eqs.(26-27) in Ref. [6].  Some systematic effects will
cancel by considering the ratios, $\zeta / \xi$ versus ${\cal
R}_W $, and  $\omega / \xi$ versus ${\cal S}_W $.

Note in this reference theory $\xi =
(|g_L|^2-
|g_R|^2)/(|g_L|^2+|g_R|^2)$.  In units of Sec. 3, $ \Gamma =
\frac{q_w}{4 \pi}
(|g_L|^2+|g_R|^2) \vert V_{tb} \vert^2 ({m_t}^2 / {m_w}^2 +1
-2 {m_w}^2 / {m_t}^2 ) $ where $g_{L,R} = \frac{1}{2} (g_V
\mp g_A )$, so in SM limit $g_L = \frac{g}{2 \sqrt2} = g_V =
-g_A$.  Note also that any $ \tilde{T}_{FS} $
violation is ``masked" since $\omega^{'} = \eta^{'} = 0 $
(i.e.
$\beta^a = \beta^b = 0 $) automatically. This ``$V$ and $A$,
$m_b
= 0$" masking mechanism could be partially the
cause
for why $T$ violation has not been
manifest in previous experiments with the lighter quarks and
leptons, even if it is not
suppressed
in the fundamental electroweak Lagrangian.

(6) The ``additional structure'' due to additional Lorentz
couplings in ${J}_{\bar b t}$ can show up
experimentally because of
its interference with the $(V-A)$ part which, we assume,
arises as predicted
by the SM.

(7) Besides model independence, a major open issue is
whether
or not there is an additional chiral coupling in the $t$
quark's
charged-current. A chiral classification of
additional structure is a natural phenomenological extension
of the
standard $SU(2)_L\ X\ U(1)$\ electroweak
symmetry. The
requirement of $\bar u(p_b )\rightarrow \bar u(p_b )\frac
12(1+\gamma _5)
$ and/or $u(k_t )\rightarrow \frac 12(1-\gamma _5)u(k_t
)$\ invariance
of the vector and axial current matrix elements $\langle b
\left| v^\mu
(0)\right| t \rangle $\ and $\langle b \left| a^\mu
(0)\right| t
\rangle $,$\ $allows only $g_L,g_{S+P},g_{S^{-}+P^{-
},}g_{+}=f_M+f_E,$and $%
\tilde g_{+}=T^{+}+T_5^{+}\ $couplings. From this $SU(2)_L$
perspective, the
relevant experimental question is what are the best limits on
such
additional couplings? Similarly, $\bar u(p_b )\rightarrow
\bar u(p_b
)\frac 12(1-\gamma _5)$ and/or $u(k_t )\rightarrow \frac
12(1+\gamma _5)\
u(k_t )$\ invariance selects the complimentary set of $%
g_R,g_{S-P},g_{S^{-}-P^{-},}g_{-}=f_M-f_E,$and $\tilde g_{-
}=T^{+}-T_5^{+}\ $%
couplings. The absence of $SU(2)_R$ couplings is simply built
into the
standard model; it is not predicted by it. So in the near
future, it will be important to ascertain the
limits on such $SU(2)_R$ couplings in $t$ quark physics.

(8) In a
separate paper  \cite{PL}, it
has been reported that at the Tevatron, percent level
statistical uncertainties are typical for
measurements of
the helicity parameters $\xi, \zeta, \sigma, \omega, \eta$.
At the LHC, several mill level
uncertainties are typical.   These are also the sensitivity
levels found for measurement of the
polarized-partial-widths, $\Gamma_{L,T}^{\pm}$, and for the
non-CKM-type CP violation parameter $r_a =\frac{|A\left( -1,-
\frac 12\right) |}{|A\left( 0,-\frac
12\right)
|}$ versus  $ r_b=\frac{|B\left( 1,\frac 12\right)
|}{|B\left(
0,\frac
12\right) |} $.
From $I_4$, see Eq.(69) below, the $\eta$
parameter($\omega$ parameter) can respectively best be
measured at the
Tevatron(LHC).  However, by the use of additional variables
(all
of $\tilde \theta_1, \tilde \phi_1, \tilde \theta_2, \tilde
\phi_2$ as in Eq.(66) ) in the stage-two step of the decay
sequences where
$W^{\pm} \rightarrow j_{\bar d,d} j_{u,\bar u}$, or  $l^{\pm}
\nu$,  we expect that these sensitivities
would then be comparable to that for the other helicity
parameters. Inclusion of additional variables should also
improve the sensitivity to the CP violation parameter
$\beta_a$ which is at $33^o$ (Tevatron), $9.4^o$ (LHC).
In regard
to effective mass-scales for new physics
exhibited by additional Lorentz couplings,
$50-70 TeV$ effective-mass scales can be probed at the
Tevatron and $110-750 TeV$ scales at the LHC.

The cleanest measurement of these parameters would presumably
be at a future $e^- e^+$ or $\mu^- \mu^+$ collider.

In Sec. 2, we introduce the necessary helicity formalism for
describing $t\rightarrow
W^{+}b\rightarrow (l^{+}\nu )b$, and $t\rightarrow
W^{+}b\rightarrow (j_{\bar
d}j_u)b$.

In Sec. 3, we list the $A(\lambda
_{W^+} ,\lambda
_b )\ $ helicity amplitudes for $t \rightarrow W^+ b $\ for
the
most general $J_{\bar b t}$ current.  Next, the helicity
parameters
are expressed in terms of a ``$(V-
A)+$\ additional
chiral coupling'' structure in the $J_{\bar b t}$
current.  Two tables display the leading-order expressions
for the
helicity parameters when the
various additional chiral couplings $(g_i/2\Lambda _i)\ $are
small relative
to the standard $V-A$\ coupling $(g_L).$\

Sec. 4 gives the inverse formulas for extracting the
contribution of the
longitudinal and transverse $W$-bosons to the polarized-
partial-
widths,
$\Gamma_{L,T}$, and to the partial-width interference-
intensities, $I_{R,I}$, from
measured values for the helicity parameters.
Expressions are also listed for extracting the phase
differences
$\beta_a$ and $\beta_a^R$ from measured values for the
helicity
parameters.

Sec. 5 gives the derivation of the full S2SC function
for the production decay sequence $ q
\overline{q},$ or $gg\rightarrow
t\overline{t}\rightarrow (W^{+}b)(W^{-}%
\overline{b}) \rightarrow (l^{+}\nu b)(l^{-}\overline{\nu
}\overline{b})$ or $
(j\overline{_d}j_ub)(j_dj\overline{_u}\overline{b}$.  Two
simpler
S2SC are then derived. Several figures show the the $\cos
\theta
_1^t,\cos
\widetilde{%
\theta _1}$ behaviour of the elements of the integrated, or
``reduced'',
composite-density-matrix $\rho
_{hh^{^{\prime }}}$.  It is this behaviour, i.e. the use of
$W$ decay-polarimetry, which is
responsible for
the enhanced sensitivity of the S2SC function $I_4$ versus
the
energy-energy spin-correlation function $I(E_{W^+},E_{W^-})$.

Sec. 6 contains some additional remarks.

\section{ THE HELICITY FORMALISM FOR \newline $t\rightarrow
W^{+}b\rightarrow (l^{+}\nu )b$, AND $t\rightarrow
W^{+}b\rightarrow (j_{\bar
d}j_u)b$:}

In the  $t$ rest frame, the
matrix element for $t \rightarrow W^{+} b$ is
\beq
\langle \theta _1^t ,\phi _1^t ,\lambda _{W^{+} } ,\lambda
_b
|\frac
12,\lambda _1\rangle =D_{\lambda _1,\mu }^{(1/2)*}(\phi
_1^t ,\theta
_1^t ,0)A\left( \lambda _{W^{+} } ,\lambda _b \right)
\eeq
where $\mu =\lambda _{W^{+} } -\lambda _b $ and $\lambda_1$
is
the $t$
helicity.  The final $W^{+}$ momentum is in the $\theta _1^t
,\phi _1^t$ direction, see Fig. 1.  For the $CP$-conjugate
process, $\bar t \rightarrow W^{-} \bar b$, in the $\bar t$
rest
frame
\beq
\langle \theta _2^t ,\phi _2^t ,\lambda _{W^{-} },\lambda
_{\bar b}|\frac 12,\lambda _2\rangle =D_{\lambda _2,\bar \mu
}^{(1/2)*}(\phi
_2^t ,\theta _2^t ,0)B\left( \lambda _{W^{-} },\lambda
_{\bar b}\right)
\eeq
with $\bar \mu =\lambda _{W^{-}}-\lambda _{\bar b}$,
$\lambda_2$ is
the $\bar t$
helicity.  Rotational invariance forbids the other $W^+$ and
$W^-$ amplitudes, so there are only two, and not three
amplitudes $A(0, -1/2), A(-1, -1/2) $ for $t \rightarrow W^+
b_L,$ etc.  An elementary, technical point[11] is that we
have
set the third Euler angle equal to zero in the big D
functions in Eqs.(8,9).  A nonzero value of the third Euler
angle would imply an (ackward) associated rotation about the
final $W^+$ momentum direction in Fig. 2.  This technical
point is important in this paper because in the spin-
correlation we exploit the azimuthal angular dependence of
the second-stage,  $W^{+} \rightarrow l^+ \nu$ or for $W^{+}
\rightarrow j_{\bar d} j_{u}$, in the decay sequences.

Fig. 2  defines the usual spherical angles
$\tilde \theta _a$, $%
\tilde \phi _a$ which specify the $j_{\bar d}$ jet (or the
$l^{+}$)
momentum in the $W^{+}
$rest frame when the boost is from the $t$ rest frame. For
the
hadronic $W^{+}$ decay mode, we use the notation that the
momentum of the charge $\frac{1}{3} e$ jet is denoted by
$j_{\bar d}$ and the momentum of the charge $\frac{2}{3} e$
jet by $j_{u}$. Likewise, Fig. 3 defines the $\tilde \theta
_b$, $%
\tilde \phi _b$ which specify the $j_{d}$ jet (or the
$l^{-}$)
momentum which occurs in the $CP$-conjugate decay sequence.

As shown in Fig. 4, we use subscripts ``$1,2$" in place of
``$a,b$" when the boost to these $W^{\pm}$ rest frames is
directly from the $(t \bar t)_{cm} $ center-of-mass frame.
Physically these angles, $\tilde \theta _a$, $%
\tilde \phi _a$ and $\tilde \theta _1$, $%
\tilde \phi _1$, are simply related by a Wigner-rotation, see
Eqs.(74,75) below.  For the CP-conjugate mode, one only needs
to change the subscripts $a \rightarrow b, 1 \rightarrow 2$.

In the $W^{+}$ rest frame, the matrix
element for $W^{+} \rightarrow l^+ \nu$ or for $W^{+}
\rightarrow j_{\bar d} j_{u}$ is [12,13]
\beq
\langle \tilde \theta _a ,\tilde \phi _a ,\lambda _{l^+}
,\lambda
_{\nu}
| 1,\lambda _{W^+} \rangle =D_{\lambda _{W^+},1 }^{1*}
 (\tilde \phi
_a ,\tilde \theta
_a ,0)c
\eeq
since $\lambda
_{\nu}= - \frac{1}{2}, \lambda
_{l^+}= \frac{1}{2}$, respectively neglecting
$(\frac{m_l}{m_W})$
corrections, neglecting $(\frac{ m_{jet} }{m_W})$
corrections.

The associated composite decay-density-matrix for
$t\rightarrow
W^{+}b\rightarrow (l^{+}\nu )b$, or for $t\rightarrow
W^{+}b\rightarrow (j_{\bar
d}j_u)b$, is
\begin{equation}
R_{\lambda _1\lambda _1^{^{\prime }}}=\sum_{\lambda
_W,\lambda
_W^{^{\prime
}}}\rho _{\lambda _1\lambda _1^{^{\prime }};\lambda _W\lambda
_W^{^{\prime
}}}(t\rightarrow W^{+}b)\rho _{\lambda _W\lambda _W^{^{\prime
}}}(W^{+}\rightarrow l^{+}\nu )
\end{equation}
where $\lambda
_W,\lambda
_W^{^{\prime
}}=0, \pm 1 $ with

$$
\rho _{\lambda _1\lambda _1^{^{\prime }};\lambda _W\lambda
_W^{^{\prime
}}}(t\rightarrow W^{+}b)=\sum_{\lambda _b=\mp 1/2}D_{\lambda
_1,\mu
}^{(1/2)*}(\phi _1^t,\theta _1^t,0)D_{\lambda _1^{^{\prime
}},\mu ^{^{\prime
}}}^{(1/2)}(\phi _1^t,\theta _1^t,0)A(\lambda _W,\lambda
_b)A(\lambda
_W^{^{\prime
}},\lambda _b)^{*}
$$
$$
\rho _{\lambda _W\lambda _W^{^{\prime }}}(W^{+}\rightarrow
l^{+}\nu
)=D_{\lambda _W,1}^{1*}(\widetilde{\phi _a},\widetilde{\theta
_a}%
,0)D_{\lambda _W^{^{\prime }},1}^1(\widetilde{\phi
_a},\widetilde{\theta _a}%
,0) |c|^2
$$
This composite decay-density-matrix can be
expressed in terms of the eight helicity parameters:
\ber
{\bf R=}\left(
\begin{array}{cc}
{\bf R}_{++} & e^{\iota \phi _1^t }
{\bf r}_{+-} \\ e^{-\iota \phi _1^t }{\bf r}_{-+} & {\bf
R}_{--}
\end{array}
\right)
\eer
The diagonal
elements  are
\beq
{\bf R}_{\pm \pm }={\bf n}_a[1\pm {\bf f}_a\cos \theta
_1^t ]\pm (1/\sqrt{%
2})\sin \theta _1^t \{\sin 2 \tilde \theta _a\
[\omega \cos \tilde \phi
_a+\eta ^{\prime }\sin \tilde \phi _a] - 2 \sin \tilde \theta
_a\
[\eta \cos \tilde \phi
_a+\omega ^{\prime }\sin \tilde \phi _a] \}
\eeq
The off-diagonal elements depend on
\ber
\begin{array}{c}
{\bf r}_{+-}=({\bf r}_{-+})^{*} \\ ={\bf n}_a{\bf f}_a\sin
\theta _1^t
+ \sqrt{2} \sin \tilde \theta _a  \{ \cos
\theta _1^t [\eta \cos \tilde
\phi _a+\omega ^{\prime }\sin \tilde \phi _a]+\iota [\eta
\sin \tilde \phi
_a-\omega ^{\prime }\cos \tilde \phi _a]\}\\
- \frac{1}{\sqrt{2} } \sin 2 \tilde \theta _a \{ \cos
\theta _1^t [\omega \cos \tilde
\phi _a+\eta ^{\prime }\sin \tilde \phi _a]+\iota [\omega
\sin \tilde \phi
_a-\eta ^{\prime }\cos \tilde \phi _a]\}
\end{array}
\eer
In Eqs.(13,14),%
\ber
\begin{array}{c}
{\bf n}_a=\frac 1{8}(5-\cos 2\tilde \theta _a-\sigma
[1+3\cos 2\tilde
\theta _a]-4 [\xi-\zeta] \cos \tilde \theta _a ) \\
{\bf n}_a{\bf f}_a=\frac 1{8}(4 [1-\sigma] \cos \tilde \theta
_a -\xi [1+3\cos
2\tilde
\theta _a]+\zeta [5-\cos 2\tilde \theta _a])
\end{array}
\eer
or equivalently%
\ber
\left(
\begin{array}{c}
{\bf n}_a \\ {\bf n}_a{\bf f}_a
\end{array}
\right) =\sin ^2\tilde \theta _a\frac{\Gamma _L^{\pm
}}{\Gamma }\pm
\frac 14(3+\cos 2\tilde \theta _a ) \frac{\Gamma _T^{\pm
}}{\Gamma
}\mp \cos \tilde \theta _a  \frac{\Gamma _T^{\mp }}{\Gamma
}
\eer

For the CP-conjugate process  $\bar t \rightarrow
W^{-} \bar b \rightarrow (l^{-} \bar\nu ) \bar b$ or $\bar t
\rightarrow W^{-} \bar b \rightarrow (j_{d}j_{\bar u}) \bar
b$
\ber
{\bf \bar R=}\left(
\begin{array}{cc}
{\bf \bar R}_{++} & e^{\iota \phi _2^t }
{\bf \bar r}_{+-} \\ e^{-\iota \phi _2^t }{\bf \bar r}_{-
+} & {\bf \bar R}%
_{--}
\end{array}
\right)
\eer
\beq
{\bf \bar R}_{\pm \pm }={\bf n}_b[1\mp {\bf f}_b\cos \theta
_2^t ]\mp (1/%
\sqrt{2})\sin \theta _2^t  \{\sin 2\tilde \theta _b [ \bar
\omega \cos \tilde
\phi _b-\bar \eta ^{\prime }\sin \tilde \phi _b] - 2 \sin
\tilde
\theta _b [ \bar \eta \cos \tilde
\phi _b-\bar \omega ^{\prime }\sin \tilde \phi _b] \}
\eeq
\ber
\begin{array}{c}
{\bf \bar r}_{+-}=({\bf \bar r}_{-+})^{*} \\ =-{\bf n}_b{\bf
f}_b\sin \theta
_2^t - \sqrt{2} \sin \tilde \theta _b  \{\cos \theta _2^t
[\bar
\eta \cos \tilde \phi _b-\bar \omega ^{\prime }\sin \tilde
\phi _b]+\iota
[\bar \eta \sin \tilde \phi _b+\bar \omega ^{\prime } \cos
\tilde \phi _b]\}\\
+ \frac{1}{\sqrt{2} }  \sin 2 \tilde \theta _b  \{\cos \theta
_2^t [\bar
\omega \cos \tilde \phi _b-\bar \eta ^{\prime }\sin \tilde
\phi _b]+\iota
[\bar \omega \sin \tilde \phi _b+\bar \eta ^{\prime } \cos
\tilde \phi _b]\}
\end{array}
\eer
\ber
\begin{array}{c}
{\bf n}_b=\frac 1{8}(5-\cos 2\tilde \theta _b-\bar \sigma
[1+3\cos
2\tilde \theta _b]-4[\bar \xi - \bar \zeta]\cos
\tilde \theta _b) \\
{\bf n}_b{\bf f}_b=\frac 1{8}(4[1 - \bar \sigma]\cos
\tilde \theta _b -\bar \xi
[1+3\cos 2\tilde \theta _b]+\bar \zeta [5-\cos
2\tilde \theta _b])
\end{array}
\eer
\ber
\left(
\begin{array}{c}
{\bf n}_b \\ {\bf n}_b{\bf f}_b
\end{array}
\right) =\sin ^2\tilde \theta _b\frac{\bar \Gamma _L^{\pm
}}{\bar \Gamma }%
\pm \frac 14(3+\cos 2\tilde \theta _b) \frac{\bar \Gamma
_T^{\pm
}}{\bar
\Gamma } \mp \cos \tilde \theta _b \frac{\bar \Gamma _T^{\mp
}}{\bar
\Gamma }
\eer

\section{ THE HELICITY PARAMETERS IN TERMS OF CHIRAL
COUPLINGS}

For \hskip1em  $t \rightarrow W^+ b$, the most general
Lorentz
coupling is
\begin{equation}
W_\mu ^{*}\bar u_{b}\left( p\right) \Gamma ^\mu
u_t \left(
k\right)
\end{equation}
where $k_t =q_w +p_b $. In (22)
\begin{eqnarray*}
\Gamma _V^\mu =g_V\gamma ^\mu +
\frac{f_M}{2\Lambda }\iota \sigma ^{\mu \nu }(k-p)_\nu   +
\frac{g_{S^{-}}}{2\Lambda }(k-p)^\mu  \\
+\frac{g_S}{2\Lambda
}(k+p)^\mu
+%
\frac{g_{T^{+}}}{2\Lambda }\iota \sigma ^{\mu \nu }(k+p)_\nu
\end{eqnarray*}
\begin{eqnarray*}
\Gamma _A^\mu =g_A\gamma ^\mu \gamma _5+
\frac{f_E}{2\Lambda }\iota \sigma ^{\mu \nu }(k-p)_\nu \gamma
_5
+
\frac{g_{P^{-}}}{2\Lambda }(k-p)^\mu \gamma
_5  \\
+\frac{g_P}{2\Lambda }%
(k+p)^\mu \gamma _5  +\frac{g_{T_5^{+}}}{2\Lambda }\iota
\sigma ^{\mu \nu
}(k+p)_\nu \gamma _5
\end{eqnarray*}
The parameter
$%
\Lambda =$ ``the effective-mass scale of new physics''.

Without additional theoretical or
experimental
inputs, it is not possible to select what is the ``best"
minimal set of couplings for
analyzing the structure of the ${ J_{\bar b t} }$ current.
There are the
``equivalence theorems" that for the
vector current, $
S\approx V+f_M,  T^{+}\approx -V+S^{-}$,
and for the axial-vector current,
$
P\approx -A+f_E,  T_5^{+}\approx A+P^{-}$.  On the
other
hand, dynamical considerations such as compositeness
would suggest searching for an additional tensorial
$g_{+}=f_M +
f_E$ coupling which would preserve $\xi =1$ but otherwise
give
non-($V-A$)-values to the $t$ helicity parameters. For
instance, $\sigma =\zeta
\neq 0.41$ and $%
\eta =\omega \neq 0.46$.

The matrix elements of the divergences of these charged-
currents are
\beq
(k-p)_\mu V^\mu =[g_V(m_t -m_b)
+
\frac{g_{S^{-}}}{2\Lambda }q^2+\frac{g_S}{2\Lambda }(m_t
^2-m_b ^2)
 +%
\frac{g_{T^{+}}}{2\Lambda }(q^2-[m_t -m_b ]^2)]\bar
u_b
u_t
\eeq
\beq
(k-p)_\mu A^\mu =[- g_A(m_b +m_t)
+
\frac{g_{P^{-}}}{2\Lambda }q^2+\frac{g_P}{2\Lambda }(m_t
^2-m_b ^2)
 +%
\frac{g_{T_5^{+}}}{2\Lambda }(q^2-[m_t+m_b ]^2)]\bar
u_b \gamma
_5u_t
\eeq
Both the weak magnetism  $\frac{f_M}{2\Lambda }$ and the weak
electricty $%
\frac{f_E}{2\Lambda }$ terms are divergenceless. On the other
hand, since $%
q^2=m_w ^2$,  even when $m_b =m_t $ there are non-
vanishing
terms due to
the couplings $S^{-},T^{+},A,P^{-},T_5^{+}$.

The modularity and simple symmetry relations\cite{PL} among
the $t
\rightarrow W^+
b$, $ \bar t \rightarrow W^- \bar b$ amplitudes
are possible
because of the phase conventions that were built
into the helicity formalism\cite{JW}.  In combining these
amplitudes with results from calculations of
similar amplitudes by diagramatic methods, care must be
exercised to insure that the same phase
conventions are being used (c.f. appendix in
[11]).

The helicity amplitudes for $t \rightarrow W^+ b_{L,R}$ for
both $(V\mp A)$ couplings and
$%
m_b $ arbitrary are for $b_L$ so $\lambda _b =-\frac
12$,%
\ber
A\left( 0,-\frac 12\right) & = & g_L
\frac{E_w +q_w }{m_w } \sqrt{m_t \left( E_b
+q_w
\right) } -g_R
\frac{E_w -q_w }{m_w } \sqrt{m_t \left( E_b -
q_w
\right) } \\ A\left( -1,-\frac 12\right) & = & g_L
\sqrt{2m_t \left( E_b +q_w \right) } -
g_R\sqrt{2m_t \left(
E_b -q_w \right) }.
\eer
and for $b _R$ so $\lambda _b =\frac 12$,%
\ber
A\left( 0,\frac 12\right) & = & -g_L
\frac{E_w -q_w }{m_w } \sqrt{m_t \left( E_b -
q_w
\right) }  +g_R
\frac{E_w +q_w }{m_w } \sqrt{m_t \left( E_b
+q_w
\right) } \\ A\left( 1,\frac 12\right) & = & -g_L
\sqrt{2m_t \left( E_b -q_w \right) }
+g_R\sqrt{2m_t \left(
E_b +q_w \right) }
\eer
Note that
$g_L,g_R$
denote the
`chirality' of the coupling and $\lambda _b =\mp \frac 12$
denote
the
handedness of $b_{L,R}$.  For $(S \pm P)$ couplings, the
additional contributions are
\ber
A(0,-\frac 12) & =g_{S+P}(
\frac{m_t }{2\Lambda })\frac{2q_w }{m_w
}\sqrt{m_t
(E_b
+q_w )}  +g_{S-P}(\frac{m_t }{2\Lambda })\frac{2q_w
}{m_w }%
\sqrt{m_t (E_b -q_w )}, \quad
A(-1,-\frac 12) & =0
\eer
\ber
A(0,\frac 12) & =g_{S+P}(
\frac{m_t }{2\Lambda })\frac{2q_w }{m_w
}\sqrt{m_t
(E_b
-q_w )}  +g_{S-P}(\frac{m_t }{2\Lambda })\frac{2q_w
}{m_w }%
\sqrt{m_t (E_b +q_w )}, \quad
A(1,\frac 12) & =0
\eer
The two types of tensorial couplings, $g_\pm = f_M \pm f_E$
and $\tilde{g}_{\pm}={g^+}_{T^+ \pm T_5^+}$,  give the
additional contributions

\begin{eqnarray*}
A\left( 0,\mp\frac 12\right)         & = & \mp g_{+} (
\frac{m_t}{2\Lambda }) \left[
\frac{E_w \mp q_w }{m_w} \sqrt{m_t \left( E_b
\pm q_w
\right) } - \frac{m_b }{m_t}
\frac{E_w \mp q_w }{m_w } \sqrt{m_t \left( E_b
\mp
q_w
\right) } \right] \\
                                             &   & \pm g_{-}
(
\frac{m_t }{2\Lambda }) \left[
 - \frac{m_b }{m_t }
\frac{E_w \pm q_w }{m_w } \sqrt{m_t \left( E_b
\pm
q_w
\right) } + \frac{E_w \pm q_w }{m_w } \sqrt{m_t
\left( E_b
\mp q_w
\right) } \right] \\
                                           &   & \mp \tilde
g_{+} (
\frac{m_t }{2\Lambda }) \left[
\frac{E_w \pm q_w }{m_w } \sqrt{m_t \left( E_b
\pm q_w
\right) } + \frac{m_b }{m_t }
\frac{E_w \mp q_w }{m_w } \sqrt{m_t \left( E_b
\mp
q_w
\right) } \right] \\
                                             &   & \pm \tilde
g_{-} (
\frac{m_t }{2\Lambda }) \left[
  \frac{m_b }{m_t }
\frac{E_w \pm q_w }{m_w } \sqrt{m_t \left( E_b
\pm
q_w
\right) } + \frac{E_w \mp q_w }{m_w } \sqrt{m_t
\left( E_b
\mp q_w
\right) } \right]
\end{eqnarray*}
\begin{eqnarray*}
A\left( \mp 1,\mp\frac 12\right) & = & \mp \sqrt{2} g_{+} (
\frac{m_t }{2\Lambda }) \left[
  \sqrt{m_t \left( E_b
\pm q_w
\right) } -  \frac{m_b }{m_t }
 \sqrt{m_t \left( E_b \mp
q_w
\right) } \right] \\
                                             &   & \pm
\sqrt{2} g_{-} (
\frac{m_t }{2\Lambda }) \left[
 - \frac{m_b }{m_t }
 \sqrt{m_t \left( E_b \pm
q_w
\right) } +  \sqrt{m_t \left( E_b
\mp q_w
\right) } \right] \\
                                  &   & \mp \sqrt{2} \tilde
g_{+} (
\frac{m_t }{2\Lambda }) \left[
 \sqrt{m_t \left( E_b
\pm q_w
\right) } + \frac{m_b }{m_t }
 \sqrt{m_t \left( E_b \mp
q_w
\right) } \right] \\
                                             &   & \pm
\sqrt{2} \tilde g_{-} (
\frac{m_t }{2\Lambda }) \left[
  \frac{m_b }{m_t }
 \sqrt{m_t \left( E_b \pm
q_w
\right) } +  \sqrt{m_t \left( E_b
\mp q_w
\right) } \right]
\end{eqnarray*}
\begin{equation}
\end{equation}

\subsection{Helicity parameters' form in terms of $g_L$
plus one \newline ``additional chiral coupling'' }

We first display the expected forms for the above helicity
parameters
for the $t \rightarrow W^+ b$ decay for
the case of a pure $V-A$ chiral coupling as in the SM. Next
we will give the form for the case of a single chiral
coupling $%
(g_i/2\Lambda _i)$\ in addition to the standard $V-A$
coupling. In this
case, we first list the formula for an arbitrarily large
additional
contribution.

In Tables 3 and 4 we list the formulas to leading order in
$g_i$ versus the standard $g_L$ coupling.
Throughout this
paper, we usually suppress the entry in the ``$i$'' subscript
on the
new-physics coupling-scale ``$\Lambda _i$'' when it is
obvious from the
context of interest.

In the case of ``multi-additional'' chiral contributions, the
general
formulas for $A(\lambda_{W^+} ,\lambda_b )$\ \ which are
listed above can be substituted into the above definitions so
as
to derive the
expression(s) for the ``multi-additional'' chiral
contributions. The $m_b / m_w, m_b / m_t$ corrections to the
following expressions can similarly be included.

{\em Pure }$V-A$ {\em coupling:}
\begin{equation}
\begin{array}{cc}
\xi =\sigma / {\cal S}_W =\zeta / {\cal S}_W = \omega / {\cal
R}_W =\eta / {\cal R}_W=1 \\ \omega ^{\prime }=
\eta ^{\prime }=0
\end{array}
\end{equation}

$V+A{\em \ also\ present:}$%
\begin{equation}
\begin{array}{cc}
\zeta / {\cal S}_W= \xi  & \omega / {\cal R}_W=\xi  \\ \sigma
/ {\cal S}_W=1  & \eta / {\cal R}_W=
1  \\ \xi =\frac{\left| g_L\right| ^2-\left| g_R\right| ^2}{%
\left| g_L\right| ^2+\left| g_R\right| ^2} & \omega ^{\prime
}=\eta ^{\prime
}=0
\end{array}
\end{equation}

$S+P\ {\em also\ present:}$%
\begin{equation}
\zeta =\sigma =\left(
\begin{array}{c}
(1-2
\frac{m_w ^2}{m_t ^2})\left| g_L\right|
^2+\frac{m_t}{\Lambda}
[1-
\frac{m_w ^2%
}{m_t ^2}]{\cal RE}(g_L^{*}g_{S+P}) \\ +\{
\frac{m_t}{2\Lambda
}[1-
\frac{m_w ^2}{%
m_t ^2}]\}^2\left| g_{S+P}\right| ^2
\end{array}
\right) / (  {\cal D}^{+} )
\end{equation}
\begin{equation}
\xi =1
\end{equation}
\begin{equation}
\begin{array}{c}
\omega =\eta =
\sqrt{2}\frac{m_w}{m_t} \left( \left| g_L\right| ^2+\frac{
m_t}{2\Lambda }[1-%
\frac{m_w ^2}{m_t ^2}]{\cal RE}(g_L^{*}g_{S+P})\right) / (
 {\cal D}^{+} ) \\ \omega
^{\prime }=\eta ^{\prime }=-\sqrt{2}\frac{m_w }{2\Lambda
}[1-\frac{m_w
^2}{m_t ^2}]{\cal IM}(g_L^{*}g_{S+P})/ ( {\cal
D}^{+} )
\end{array}
\end{equation}
where
$$
{\cal D}^{+ }=(1+2\frac{m_w ^2}{m_t ^2})\left| g_L\right|
^2+\frac{m_t}{\Lambda} [1-%
\frac{m_w ^2}{m_t ^2}]{\cal RE}(g_L^{*}g_{S+P})+\{ \frac{
m_t}{2\Lambda } [1-\frac{%
m_w ^2}{m_t ^2}]\}^2\left| g_{S+P}\right| ^2
$$

$S-P\ {\em also\ present:}$%
\begin{equation}
\zeta ,\sigma =\left( (1-2\frac{m_w ^2}{m_t ^2})\left|
g_L\right| ^2\mp
\{ \frac{m_t}{2\Lambda} [1-(\frac{m_w ^2}{m_t ^2})]\}^2\left|
g_{S-P}\right|
^2\right) / (  {\cal D}^{-} )
\end{equation}
where the upper(lower) sign on the ``rhs'' goes with the
first(second) entry
on the ``lhs.''
\begin{equation}
\xi =\left( (1+2\frac{m_w ^2}{m_t ^2})\left|
g_L\right| ^2 -
\{ \frac{m_t}{2\Lambda} [1-(\frac{m_w ^2}{m_t ^2})]\}^2\left|
g_{S-P}\right|
^2\right) / (  {\cal D}^{-} )
\end{equation}
\begin{equation}
\omega =\eta =\sqrt{2} \frac{m_w}{m_t} \left| g_L\right| ^2/(
 {\cal D}^{-} ),\ \
\omega ^{\prime }=\eta ^{\prime }=0
\end{equation}
where
$$
{\cal D}^{-}=(1+2\frac{m_w ^2}{m_t ^2})\left| g_L\right|
^2+\{ \frac{m_t}{2\Lambda} [1-\frac{m_w ^2}{m_t
^2}]\}^2\left| g_{S-
P}\right| ^2
$$

$f_M+f_E\ {\em also\ present:}$

For this case we write the coupling constant of the sum of
the weak
magnetism and the weak electricity couplings as
$$
g_{+}=f_M+f_E
$$
In this notation,
\begin{equation}
\zeta =\sigma =\left(
\begin{array}{c}
(1-2
\frac{m_w ^2}{m_t ^2})\left| g_L\right| ^2+\frac{m_w
^2}{m_t \Lambda }{\cal %
RE}(g_L^{*}g_{+}) \\ +\frac{m_w ^2}{4\Lambda ^2}[-
2+\frac{m_w ^2}{m_t ^2}%
]\left| g_{+}\right| ^2
\end{array}
\right) / ( {\cal D_T}^{+} )
\end{equation}
$$
\xi =1
$$
\begin{equation}
\begin{array}{c}
\omega =\eta =
\sqrt{2} \frac{m_w}{m_t} \left( \left| g_L\right| ^2- \frac{
m_t}{2\Lambda} [1+%
\frac{m_w ^2}{m_t ^2}]{\cal RE}(g_L^{*}g_{+})+\frac{m_w
^2}{4\Lambda ^2}%
\left| g_{+}\right| ^2\right) / (  {\cal
D_T}^{+} )  \\ \omega ^{\prime }=\eta
^{\prime }=-\frac{m_w }{\sqrt{2}\Lambda }[1-\frac{m_w
^2}{m_t ^2}]{\cal IM%
}(g_L^{*}g_{+})/ ( {\cal D_T}^{+} )
\end{array}
\end{equation}
where
$$
{\cal D_T}^{+}=(1+2\frac{m_w ^2}{m_t ^2})\left| g_L\right|
^2-3\frac{m_w ^2%
}{m_t \Lambda }{\cal RE}(g_L^{*}g_{+})+\frac{m_w ^2}{4\Lambda
^2}[2+\frac{%
m_w ^2}{m_t ^2}]\left| g_{+}\right| ^2
$$
$f_M-f_E\ {\em also\ present:}$

Similarly, we write the coupling constant of the difference
of the weak
magnetism and the weak electricity couplings as
$$
g_{-}=f_M-f_E
$$
and so,
\begin{equation}
\zeta ,\sigma =\left( (1-2\frac{m_w ^2}{m_t ^2})\left|
g_L\right| ^2\pm
\frac{m_w ^2}{4\Lambda ^2}\left| g_{-}\right| ^2\right) /
(  {\cal D}_T^{-} )
\end{equation}
where the upper(lower) sign on the ``rhs'' goes with the
first(second) entry
on the ``lhs.''Also,
\begin{equation}
\xi =\left( (1+2\frac{m_w ^2}{m_t ^2})\left| g_L\right| ^2-
3\frac{m_w ^2}{%
4\Lambda ^2}\left| g_{-}\right| ^2\right) /  {\cal D}_T^{-}
\end{equation}
\begin{equation}
\omega ,\eta =\sqrt{2} \frac{m_w}{m_t} \left( \left|
g_L\right|
^2\mp \frac{%
m_w ^2}{4\Lambda ^2}\left| g_{-}\right| ^2\right) / (
 {\cal D}_T^{-} ) ,\ \
\omega ^{\prime }=\eta ^{\prime }=0\
\end{equation}
Here%
$$
{\cal D}_T^{-}=(1+2\frac{m_w ^2}{m_t ^2})\left| g_L\right|
^2+3\frac{m_w
^2}{4\Lambda ^2}\left| g_{-}\right| ^2
$$

$T^{+}+T_5^{+}\ {\em also\ present:}$

We let
$$
\tilde g_{+}=g_{T+T_5}^{+}
$$
In this notation,
\begin{equation}
\zeta =\sigma =\xi=1
\end{equation}
Also
\begin{equation}
\omega =\eta =1 ;\ \ \omega ^{\prime }=\eta ^{\prime
}=0
\end{equation}
A single additional \ $\tilde g_{+}={g^{+}}_{T^{+}+T_5^{+}}\
$coupling does not
change the values from that of the pure $V-A$\ coupling.

$T^{+}-T_5^{+}\ {\em also\ present:}$

\begin{quotation}
We let
$$
\tilde g_{-}=g_{T-T_5}^{+}
$$
and so,
\begin{equation}
\zeta = \xi ,\ \ \sigma =1
\end{equation}
\begin{equation}
\xi =\frac{\left| g_L\right| ^2-\left| \frac{m_t \tilde g_{-
}}{2\Lambda }%
\right| ^2}{\left| g_L\right| ^2+\left| \frac{m_t \tilde g_{-
}}{2\Lambda }%
\right| ^2}
\end{equation}
\begin{equation}
\omega = \xi ,\ \ \eta =1,\ \omega ^{\prime }=\eta
^{\prime }=0\
\end{equation}
\end{quotation}
A single additional \ $\tilde g_{-}={g^{+}}_{T^{+}-T_5^{+}}\
$coupling is
equivalent to a single additional $V+A$\ coupling, except for
the
interpretation of their respective chirality parameters.

\subsection{ Helicity parameters to leading-order in one
\newline
``additional
chiral coupling''}

In Table 3 for the $V+A$\ and for the $S\mp P$\ couplings, we
list the
``expanded forms'' of the above expressions to leading-order
in
a
single additional chiral coupling $(g_i/2\Lambda _i)$ versus
the standard $V-A$\ coupling $(g_L)$. Similarly,
in Table 4 is
listed the formulas for the additional tensorial couplings.
The tensorial
couplings include the sum and difference of the weak
magnetism and
electricity couplings, $g_{\pm }=f_M\pm f_E$, which involve
the momentum
difference $q_w =k_t -p_b $. The alternative
tensorial couplings $%
\tilde g_{\pm }={g^{+}}_{T^{+}\pm T_5^{+}}$ instead involve
$k_t +p_b $.  In application[6] of $I_4$ to determine limits
on a pure $IM( g_+ )$, as in [6], since $RE( {g_L}^* g_+ ) =
0$, the
additional terms in Table 4 going as $ |g_{+} |^2 $ can be
used; for other than pure $IM( g_+ )$, one should work
directly from the above expressions in the text.  This remark
also applies for determination of limits for a pure $ IM(
g_{S+P} ) $ from Table 3.

Notice that, except for the following coefficients, the
formulas tablulated in
these two tables are short and simple. As above we usually
suppress the
entry in the ``$i$'' subscript on ``$\Lambda _i$.'' For
\newline Table 3 these
coefficients are
\begin{equation}
\begin{array}{cc}
a=\frac{4m_w ^2}{m_t \Lambda }\frac{(1-\frac{m_w
^2}{m_t ^2})}{(1-4\frac{%
m_w ^4}{m_t ^4})} & d= \frac{m_t}{4\Lambda} (1-
\frac{m_w ^2}{m_t ^2})\frac{(1-2\frac{m_w
^2}{m_t ^2})}{(1+2\frac{m_w ^2}{%
m_t ^2})} \\ b=\frac{m_t ^2}{2\Lambda ^2}\frac{(1-\frac{m_w
^2}{m_t ^2})^2}{(1-4%
\frac{m_w ^4}{m_t ^4})} & e=
\frac{m_t ^2}{4\Lambda ^2}\frac{(1-\frac{m_w
^2}{m_t ^2})^2}{(1+2\frac{m_w ^2%
}{m_t ^2})} \\ c=\frac{m_w ^2}{\Lambda ^2}\frac{(1-
\frac{m_w ^2}{m_t ^2})^2}{%
(1-4\frac{m_w ^4}{m_t ^4})} & f= \frac{m_t}{2\Lambda} (1-
\frac{m_w ^2}{m_t ^2})
\end{array}
\end{equation}
The additional coefficients for Table 4 are
\begin{equation}
\begin{array}{cc}
g=\frac{2m_w ^2}{m_t \Lambda }\frac{(1-4\frac{m_w
^2}{m_t ^2})}{(1-4\frac{%
m_w ^4}{m_t ^4})} & l=
\frac{m_t (1+9\frac{m_w ^2}{m_t ^2}+2\frac{m_w
^4}{m_t ^4})}{2\Lambda (1+2\frac{%
m_w ^2}{m_t ^2})} \\ h=\frac{m_w ^2}{2\Lambda ^2}\frac{(1-
4\frac{m_w ^2%
}{m_t ^2})}{(1-4\frac{m_w ^4}{m_t ^4})} & n=
\frac{m_w ^2(2+\frac{m_w ^2}{m_t ^2})}{2\Lambda
^2(1+2\frac{m_w ^2}{m_t ^2%
})} \\ j=\frac{m_w ^2}{\Lambda ^2}\frac{(1-\frac{m_w
^2}{m_t ^2})}{(1-4%
\frac{m_w ^4}{m_t ^4})} & o=
\frac{m_w ^2(1-\frac{m_w ^2}{m_t ^2})}{2\Lambda
^2(1+2\frac{m_w ^2}{m_t ^2%
})} \\ k=\frac{3m_w ^2}{2\Lambda ^2(1+2\frac{m_w
^2}{m_t ^2})} & u= \frac{m_w ^2}{\Lambda ^2}\frac{(1-
\frac{m_w
^4}{m_t ^4})}{(1-4%
\frac{m_w ^4}{m_t ^4})}
\end{array}
\end{equation}
Notice that ${\cal O}(1/\Lambda )$ coefficients occur in the
case of an
interference with the $g_L$coupling, and that otherwise
${\cal
O}(1/\Lambda
^2)$ coefficients occur.

When the experimental precision is sensitive to effects
associated with the finite width $\sim 2.07 GeV$ of the W-
boson, then a smearing over this width and a more
sophisticated treatment of these coefficients will be
warranted. Numerically, for $m_t= 175 GeV, m_w= 80.36 GeV,
m_b = 4.5 GeV$ these coefficients are:
\begin{equation}
\begin{array}{cc}
a \Lambda = 141.6; b \Lambda^2 = 11,600;c \Lambda^2 = 4,890;
d \Lambda = 14.05; e \Lambda^2 = 3,354; f
\Lambda = 69.05;
\nonumber \\
g \Lambda = 14.07; h \Lambda^2 = 615.4; j
\Lambda^2 = 6,197; k \Lambda^2 = 6,812; l \Lambda = 183.8; n
\Lambda^2 = 5,020;
\nonumber \\
o \Lambda^2 = 1,792;u \Lambda^2 = 7,503
\end{array}
\end{equation}

In comparing the entries in these two tables, notice that
(i) a single
additional \ $\tilde g_{+}={g^{+}}_{T^{+}+T_5^{+}}\ $
coupling does not change the values from that of the
pure $V-A$\ coupling, and that (ii) a
single additional \ $\tilde g_{-}={g^{+}}_{T^{+}-T_5^{+}}\
$coupling is
equivalent to a single additional $V+A$\ coupling, except for
the
interpretation of their respective chirality parameters.
This follows as a consequence of the above ``equivalence
theorems'  and the absence
of contributions from the $S^-$ and $P^-$ couplings when the
$W^+$ is on-shell.
We have displayed this equivalence in Table 4 to emphasize
that while an assumed total absence of $\tilde
g_{\pm}$ couplings in $t \rightarrow W^+ b$ decay might be
supported by the
weaker test
of the experimental/theoretical normalization of the decay
rate ( i.e. the canonical universality test ),
empirical $V-A$ $(V+A)$ values of the helicity parameters
shown in these tables will not imply the absence of $\tilde
g_+$ ($\tilde g_-$) couplings.

\section{TESTS FOR ``NEW PHYSCS"}

In context of the helicity parameters, this topic in
discussed is a separate paper\cite{PL}.  Here we include some
useful formulas that were omitted in that discussion.

The contribution of the longitudinal($L$) and transverse($T$)
$W$-amplitudes in the decay
process is projected out by the simple formulas:
\begin{eqnarray*}
I_{
{\cal R}}^{b_L,b_R}\equiv \frac 12(I_{{\cal R}}^{+}\pm
I_{{\cal R}%
}^{-})=|A(0,\mp \frac 12)||A(\mp 1,\mp \frac 12)|\cos \beta
_a^{L,R} =\frac
\Gamma 2({\eta} \pm {\omega} )
\end{eqnarray*}
\begin{eqnarray*}
I_{
{\cal I}}^{b_L,b_R}\equiv \frac 12(I_{{\cal I}}^{+}\pm
I_{{\cal I}%
}^{-})=|A(0,\mp \frac 12)||A(\mp 1,\mp \frac 12)|\sin \beta
_a^{L,R} =\frac
\Gamma 2({\eta^{\prime }} \pm {\omega^{\prime }}
)
\end{eqnarray*}
\begin{eqnarray*}
\Gamma _L^{b_L,b_R}\equiv \frac 12(I_L^{+}\pm I_L^{-
})=|A(0,\mp \frac
12)|^2
=\frac \Gamma 4(1+{\sigma} \pm \xi \pm {\zeta} )
\end{eqnarray*}
\begin{eqnarray}
\Gamma _T^{b_L,b_R}\equiv \frac 12(I_T^{+}\pm I_T^{-
})=|A(\mp 1,\mp
\frac 12)|^2
=\frac \Gamma 4(1-{\sigma} \pm \xi \mp {\zeta} )
\end{eqnarray}
In the first line, $\beta _a^L=\beta _a$. Unitarity, requires
the two right-triangle
relations
\ber
(I_{
{\cal R}}^{b_L})^2+(I_{{\cal I}}^{b_L})^2=\Gamma
_L^{b_L}\Gamma
_T^{b_L} \\
(I_{{\cal R}}^{b_R})^2+(I_{{\cal I}}^{b_R})^2=\Gamma
_L^{b_R}\Gamma _T^{b_R}.
\eer
It is important to determine directly from experiment
whether or the $W_L$ and $W_T$ partial widths are anomalous
in nature
versus the standard $(V-A)$ predictions.  They
might have distinct
dynamical differences versus the SM predictions if
electroweak dynamical symmetry
breaking(DSB) occurs in nature.

By unitarity and the assumption that only the
minimal helicity amplitudes are needed, one can easily derive
expressions for measuring the phase differences between the
helicity amplitudes.
In the case of both $b_{L}$ and $b_{R}$ couplings,
there is
\ber
\begin{array}{c}
\cos \beta _a=\frac{I_{{\cal R}}^{b_L}}{\sqrt{\Gamma
_L^{b_L} \Gamma
_T^{b_L}} } \\
=\frac{2
(\omega +\eta )}{\sqrt{(1+\xi )^2-(\sigma
+\zeta )^2}}
\end{array}
\eer
and for the $b_R$ phase
difference,
\ber
\begin{array}{c}
\cos \beta _a^R=\frac{I_{{\cal R}}^{b_R}}{\sqrt{\Gamma
_L^{b_R} \Gamma
_T^{b_R}} } \\ =\frac{2 (\eta -
\omega )}{\sqrt{%
(1-\xi )^2-(\sigma -\zeta )^2}}
\end{array}
\eer
Also
\ber
\begin{array}{c}
\sin \beta _a=\frac{I_{{\cal I}}^{b_L}}{\sqrt{\Gamma
_L^{b_L} \Gamma
_T^{b_L}} }\\
=\frac{2
(\omega ^{^{\prime }}+\eta ^{^{\prime }})}{\sqrt{(1+\xi )^2-
(\sigma +\zeta )^2}}
\end{array}
\eer
with
\ber
\begin{array}{c}
\sin \beta _a^R=\frac{I_{{\cal I}}^{b_R}}{\sqrt{\Gamma
_L^{b_R} \Gamma
_T^{b_R}} } \\ =\frac{2 (\eta^{'} -
\omega^{'} )}{\sqrt{%
(1-\xi )^2-(\sigma -\zeta )^2}}
\end{array}
\eer
Measurement of $\beta _a \neq 0 (\beta _b \neq 0)$implies a
violation of $T$
invariance in $t \rightarrow W^+ b ( \bar t \rightarrow W^-
\bar b )$ or the presence of an unexpected final-state
interaction
between the $b $ and $W^{+}$. Because of the further
assumption of
no-unusual-final-state-interactions,  one is
actually
testing for
$\tilde{T_{FS}}$ invariance. Canonical $T$ invariance relates
$t \rightarrow
W^+ b $ and the actual time-reversed process  $W^+ b
\rightarrow t$ which is not directly accessible by
present
experiments. Equivalent to the two right-triangle relations
are two expressions involving the helicity parameters:
\begin{equation}
({\eta} \pm {\omega} )^2+({\eta^{\prime }}
\pm
{\omega^{\prime }} )^2=\frac
14[(1 \pm \xi )^2-( {\sigma} \pm {\zeta} )^2].
\end{equation}
Fig. 5 displays a simple test of $\tilde T_{FS}$ invariance
using the
first relation.  With forseeable experimental precisions, the
second relation appears unlikely to be tested in the near
future.

\section{STAGE-TWO SPIN-CORRELATION FUNCTIONS}

For $t \bar t$ production at hadron colliders, a simple
consequence of the QM-factorization structure of the parton
model is that there are incident
parton longitudinal beams characterized by the Feynman $x_1$
and
$x_2$ momentum fractions instead of the known $p$ and $\bar
p (p)$
momenta. This momentum uncertainty must therefore be smeared
over in application of the following S2SC functions and in
determination\cite{PL} of the associated sensitivities for
measurement of the above helicity parameters.

\subsection{The full S2SC function:}

We consider the production-decay sequence
\begin{equation}
\begin{array}{c}
q
\overline{q}, {\textstyle or } gg\rightarrow
t\overline{t}\rightarrow (W^{+}b)(W^{-}%
\overline{b}) \\ \rightarrow (l^{+}\nu b)(l^{-
}\overline{\nu}\overline{b})%
{\textstyle or
}(j\overline{_d}j_ub)(j_dj\overline{_u}\overline{b})
\end{array}
\end{equation}

The general angular distribution in the $(t\bar t)_{cm}$ is
\begin{equation}
\begin{array}{c}
I(\Theta _B,\Phi _B;\theta _1^t,\phi _1^t;
\widetilde{\theta _a},\widetilde{\phi _a};\theta _2^t,\phi
_2^t;\widetilde{%
\theta _b},\widetilde{\phi _b})=\sum_{\lambda _1\lambda
_2\lambda
_1^{^{\prime }}\lambda _2^{^{\prime }}}\rho _{\lambda
_1\lambda _2;\lambda
_1^{^{\prime }}\lambda _2^{^{\prime }}}^{prod}(\Theta _B,\Phi
_B) \\ \times
R_{\lambda _1\lambda _1^{^{\prime }}}(t\rightarrow
W^{+}b\rightarrow \ldots )%
\overline{R_{\lambda _2\lambda _2^{^{\prime }}}}(\bar
t\rightarrow W^{-}\bar
b\rightarrow \ldots )
\end{array}
\end{equation}
where the composite decay-density-matrix $R_{\lambda
_1\lambda _1^{^{\prime
}}}$ for $t\rightarrow W^{+}b\rightarrow \ldots $is given by
Eq.(12), and
that $R_{\lambda _2\lambda _2^{^{\prime }}}$ for $\bar
t\rightarrow
W^{-}\bar b\rightarrow \ldots $ is given by Eq.(17). The
angles $\Theta
_B,\Phi _B$ give[11,12] the direction of the incident parton
beam,
i.e. the $q$
momentum or the gluon's momentum, arising from the incident
$p$ in the $%
p\bar p$, or $pp\rightarrow t\bar tX$ production process.
With Eq.(62) there
is an associated differential counting rate
\begin{equation}
\begin{array}{c}
dN=I(\Theta _B,\Phi _B;\ldots )d(\cos \Theta _B)d\Phi
_Bd(\cos \theta
_1^t)d\phi _1^t \\
d(\cos \widetilde{\theta _a})d\widetilde{\phi _a}d(\cos
\theta _2^t)d\phi
_2^td(\cos \widetilde{\theta _b})d\widetilde{\phi _b}
\end{array}
\end{equation}
where, for full phase space, the cosine of each polar angle
ranges from -1
to 1, and each azimuthal angle ranges from 0 to 2$\pi .$

Each term in Eq.(62) can depend on the angle between the $t$
and $\bar t$ decay
planes
\begin{equation}
\phi =\phi _1^t+\phi _2^t
\end{equation}
and on the angular difference
\begin{equation}
\Phi _R=\Phi _B-\phi _1^t
\end{equation}
So, we treat $\Phi _B,\Phi _R,\phi $ as the azimuthal
variables. We
integrate out $\Phi _R$. The resulting full S2SC function is
relatively
simple:
\begin{equation}
\begin{array}{c}
I(\Theta _B,\Phi _B;\phi ;\theta _1^t,
\widetilde{\theta _a},\widetilde{\phi _a};\theta
_2^t,\widetilde{\theta _b},%
\widetilde{\phi _b})=\sum_{h_1h_2}\{\rho
_{h_1h_2,h_1h_2}^{prod}R_{h_1h_1}%
\overline{R_{h_2h_2}} \\ +(\rho _{++,--}^{prod}r_{+-
}\overline{r_{+-}}+\rho
_{--,++}^{prod}r_{-+}\overline{r_{-+}})\cos \phi +i(\rho
_{++,--}^{prod}r_{+-}\overline{r_{+-}}-\rho _{--
,++}^{prod}r_{-+}\overline{%
r_{-+}})\sin \phi \}
\end{array}
\end{equation}
where $\rho _{h_1h_2,h_1h_2}^{prod}(\Theta _B,\Phi _B)$ still
depends on $%
\Theta _B,\Phi _B$ and the composite density matrix elements
are given
above. The $\theta _1^t$angular dependence can be replaced by
the $W^{+}$
energy in the the $(t\bar t)_{cm}$ and similarly $\theta
_2^t$by the $W^{-}$
energy[12]. The $\sin \phi $ dependence is the well-known
test
for $CP$%
-violation in the production process[13,5].

\subsection{Two simpler S2SC functions:}

We next integrate out some of the variables to obtain
simpler S2SC
functions. First[11], we transform to the variables of Fig. 4
and then
integrate out the two aximuthal angles $\widetilde{\phi
_{1,2}}$. This gives
a five variable S2SC with respect to the final decay
products:
\begin{equation}
\begin{array}{c}
I(\phi ;\theta _1^t,
\widetilde{\theta _1},;\theta _2^t,\widetilde{\theta
_2})=\sum_{h_1h_2}\{%
\rho _{h_1h_2,h_1h_2}^{prod}R_{h_1h_1}\overline{R_{h_2h_2}}
\\ +2\cos \phi
{\cal RE}(\rho _{++,--}^{prod}\rho _{+-}\overline{\rho _{+-
}})-2\sin \phi
{\cal IM}(\rho _{++,--}^{prod}\rho _{+-}\overline{\rho _{+-
}})
\end{array}
\end{equation}
The $\sin \phi {\cal \ }$term will vanish if both $CP$
invariance holds in $%
(t\bar t)$ production and $\beta _a=\beta _b=0$ in $t$ and
$\bar t$
decays.

Diagonal $\rho _{\pm \pm }$ and off-diagonal $\rho _{\pm \mp
}$ appear here
to describe the decay  sequence $t\rightarrow
W^{+}b\rightarrow l^{+}\nu b$,
or $j\overline{_d}j_ub$. The $CP$-conjugate sequences are
described by $%
\overline{\rho _{\pm \pm }},\overline{\rho _{\pm \mp }}$.
These integrated,
composite density matrix elements are defined by
\begin{equation}
\begin{array}{c}
\rho _{h_1h_1}\equiv \frac 1{2\pi }\int_0^{2\pi }d\phi
_1R_{h_1h_1}/|A(0,-\frac 12)|^2 \\
\overline{\rho _{h_2h_2}}\equiv \frac 1{2\pi }\int_0^{2\pi
}d\phi _1%
\overline{R_{h_2h_2}}/|B(0,\frac 12)|^2 \\ =\rho _{-h_2-h_2}(
{\textstyle subscripts } 1\rightarrow 2,a\rightarrow b) \\
\rho
_{+-
}=(\rho
_{-+})^{*}\equiv \frac 1{2\pi }\int_0^{2\pi }d\phi _1r_{+-
}/|A(0,-\frac
12)|^2 \\
\overline{\rho _{+-}}=(\overline{\rho _{-+}})^{*}\equiv \frac
1{2\pi
}\int_0^{2\pi }d\phi _1\overline{r_{+-}}/|B(0,\frac 12)|^2 \\
=-\rho _{+-}(
{\textstyle subscripts} 1\rightarrow 2,a\rightarrow b,\beta
_a\rightarrow \beta _b)
\end{array}
\end{equation}
where the last lines for the CP conjugate ones shows useful
CP substitution
rules.

By integrating out the angle $\phi $ between the $t$ and
$\bar t$ decay
planes, a simple four-variable S2SC function is obtained

\begin{equation}
\begin{array}{c}
I(E_{W^{+}},E_{W^{-}},
\widetilde{\theta _1},\widetilde{\theta
_2})= \sum_{h_1,h_2} \{\rho
_{h_1h_2,h_1h_2}^{prod}\rho _{h_1h_1}\overline{\rho
_{h_2h_2}}\} \\
= \sum_{i} \{\rho _{+-}(q_i
\overline{q_i}\rightarrow t\bar t)^{prod}[\rho
_{++}\overline{\rho _{--}}%
+\rho _{--}\overline{\rho _{++}}] \\ +\rho
_{++}(gg\rightarrow t\bar
t)^{prod}[\rho _{++}\overline{\rho _{++}}+\rho _{--
}\overline{\rho _{--}}]\}
\end{array}
\end{equation}
where the sum is over the quarks and gluons in the incident
$p \bar
p$ or $pp$.
In the second line we have assumed $CP$ invariance in the
production
processes.

The simplest kinematic measurement of the above helicity
parameters at the Tevatron and at the LHC would be through
purely
hadronic top decay modes.  CDF has reported[14]
observation of
such decays.  In this case the $(t \bar t)_{cm}$ frame is
accessible and the above $I_4$ can be used.
In a separate paper[6] we have reported that the associated
statistical sensitivities to the helicity parameters are at
the percent level for measurements at the Tevatron, and at
the several mill level for at the LHC.  Fig. 6 shows the net
$E_{W^+}, E_{W^-}$ dependence of Eq.(69).

\subsection{Integrated composite decay-density-matrix
elements:}

In (69), the composite decay-density-matrix elements are
simply
the
decay
probability for a $t_1$ with
helicity $\frac h2$ to decay $t \rightarrow W^{+} b$ followed
by  $W^{+}\rightarrow j_{\bar
d}j_u$, or $%
W^{+}\rightarrow l^{+}\nu $  since
\newline
$d{\textstyle N} / d\left( \cos \theta _1^t \right) d\left(
\cos
\tilde
\theta _1\right) =\rho _{hh}\left( \theta _1^t ,\tilde
\theta
_1\right)  $ and for the decay of the $\bar t_2$ with
helicity $\frac h2$, $\bar \rho
_{hh}=$
\newline
$\rho _{-h,-
h} (
 1 \rightarrow 2,
{\textstyle add bars}
)$.
For $t_1$ with
helicity $\frac h2$
\beq
\rho _{hh}= \rho_o+h \rho_c \cos \theta _1^t +h \rho_s \sin
\theta _1^t
\eeq
where
\ber
\rho_o =\frac 18 \{6- 2 \cos ^2\omega
_1\cos^2\tilde \theta
_1- \sin ^2\omega _1\sin ^2\tilde \theta _1
\nonumber \\
+\sigma [ 2- 6 \cos ^2\omega
_1\cos^2\tilde \theta
_1- 3 \sin ^2\omega _1\sin ^2\tilde \theta _1 ]
-4(\xi-\zeta) \cos \omega
_1 \cos \tilde \theta
_1 \}\\
\rho_c =\frac 18 \{\zeta [6- 2 \cos ^2\omega
_1\cos^2\tilde \theta
_1- \sin ^2\omega _1\sin ^2\tilde \theta _1]
\nonumber \\
+\xi [ 2- 6 \cos ^2\omega
_1\cos^2\tilde \theta
_1- 3 \sin ^2\omega _1\sin ^2\tilde \theta _1 ]
+4(1-\sigma) \cos \omega
_1 \cos \tilde \theta
_1 \}\\
\rho_s =\frac{1}{\sqrt{2}} \{\frac{1}{2} \omega \sin 2\omega
_1 [ \sin ^2\tilde \theta _1 -  2 \cos^2\tilde \theta
_1]
+2 \eta \sin \omega
_1 \cos \tilde \theta
_1 \}
\eer
with the
Wigner
rotation angle $\omega_1 =
\omega_1(E_{W^+})$.
The rotation by
$\omega_1$ is about the implicit $y_a$ axis in Fig. 2. It is
given
by[11]
\beq
 \sin \omega_1 = m_W \beta \gamma \sin \theta_1^t / p_1
\eeq
\beq
\cos \omega_1 = \frac{E_{cm} (m_t^2 - m_W^2 + [ m_t^2 +
m_W^2]
\beta \cos \theta_1^t ) }{4m_t^2 p_1}
\eeq
where $p_1=$ the
magnitude of the $W^+$ momentum in the $(t \bar t)_{cm}$
frame
and $\gamma,\beta$ describe the boost from the $(t \bar
t)_{cm}$
frame to
the $t_1$ rest frame [$\gamma= E_{cm}/(2m_t)$ with $E_{cm}=$
total energy of $t \bar t$, in $(t \bar t)_{cm}$].

Note that the $\rho_s$ term depends only
on the $W_L - W_T$ interference intensities, whereas the
$\rho_o$ and $\rho_c$ terms only depend on the polarized-
partial-widths, specifically
\ber
\rho_{o,c} =\frac 12 [ 2- 2 \cos ^2\omega
_1\cos^2\tilde \theta
_1- \sin ^2\omega _1\sin ^2\tilde \theta _1 ]
\frac{\Gamma_L^{\pm} } {\Gamma}
\nonumber \\
\pm \frac 14 [ 2+ 2 \cos ^2\omega
_1\cos^2\tilde \theta
_1+ \sin ^2\omega _1\sin ^2\tilde \theta _1 ]
\frac{\Gamma_T^{\pm} } {\Gamma}
\mp \cos \omega
_1 \cos \tilde \theta
_1 \frac{\Gamma_T^{\mp} } {\Gamma}
\eer
with $\bar \rho_{o,c} = \rho_{o,c}$  ( $1 \rightarrow 2,
{\textstyle add bars} $
).

For the off-diagonal elements, the analogous expression is
\begin{equation}
\begin{array}{c}
\rho _{+-}=\rho _c\sin \theta _1^t-
\sqrt{2}(\eta \cos \theta _1^t-i\omega ^{\prime} )\sin \omega
_1\cos
\widetilde{\theta _1} \\
+ \frac{1}{2 \sqrt{2} } (\omega \cos \theta _1^t-i\eta
^{\prime}) \sin 2\omega _1[2\cos ^2 \widetilde{\theta _1} -
\sin ^2 \widetilde{\theta _1} ]
\end{array}
\end{equation}

Figures 7-14 show the $\cos \theta _1^t,\cos
\widetilde{%
\theta _1}$ behaviour of the elements of these integrated, or
``reduced'',
composite-density-matrix $\rho
_{hh^{^{\prime }}}$ assuming the $(V-A)$ values of Table 1
for the helicity parameters. These figures also show the
dependence as the
total
center-of-mass energy $E_{cm}$ is changed. Fig. 7 is for
$\rho _{++}$
and $E_{cm}=380$ $GeV$.  The next one, Fig. 8, is for
$E_{cm}=450$
$GeV$. This dependence on $\cos \theta _1^t,\cos
\widetilde{%
\theta _1}$, i.e. the use of W decay-polarimetry,  is the
reason for the greater sensitivity of the
S2SC
function, $I_4$, than the simpler energy-energy spin-
correlation
function $I(E_{W^+},E_{W^-})$, see Sec. 6.  Figs. 9-10 show
the
behaviour of $\rho _{--}$.  The behaviours of the real and
imaginary
parts of the off-diagonal elements $\rho _{+-}$ are shown in
Figs.
11-14.  Note that to display the imaginary part with an
arbitrarily fixed overall normalization, we have set
$\omega^{\prime
}= \eta^{\prime}$ $ =1$ in Eq.(77) since in the SM the
relative phase $\beta_a^R = 0$.

If the $(V-A)$ values for the helicity parameters are
empirically found to be only approximately correct, then the
details of the dependence of $\rho
_{hh^{^{\prime }}}$ on  $\cos \theta _1^t,\cos
\widetilde{%
\theta _1},$ and $E_{cm}$ will differ but, nevertheless, the
analyzaing power of $\rho
_{hh^{^{\prime }}}$ and of ${\bf R}$ of Eq.(12) should remain
large at both the Tevatron and the LHC.

\subsection{Production density matrix elements:}

The production density matrix elements for $gg\rightarrow
t\bar t$ are
calculated by the methods in [15,12]. In the usual
helicity phase
conventions, we obtain
\begin{equation}
\begin{array}{c}
\rho _{++}(gg\rightarrow t\bar t)=\rho _{++,++}=\rho _{--,--}
\\
=\frac{m_t^2}{96E_t^2}[\frac{s^2}{(m_t^2-t)^2(m_t^2-
u)^2}][7+9\frac{p_t^2}{%
E_t^2}\cos ^2\theta _t][1+\frac{p_t^2}{E_t^2}(1+\sin ^4\theta
_t)]
\end{array}
\end{equation}
\begin{equation}
\begin{array}{c}
\rho _{+-}(gg\rightarrow t\bar t)=\rho _{+-,+-}=\rho _{-+,-+}
\\
=\frac{p_t^2}{96E_t^2}[\frac{s^2}{(m_t^2-t)^2(m_t^2-
u)^2}][7+9\frac{p_t^2}{%
E_t^2}\cos ^2\theta _t]\sin ^2\theta _t(1+\cos ^2\theta _t)
\end{array}
\end{equation}
where $E_t$ is the energy of the produced $t$ quark with
momentum of
magnitude $p_t$ at angle $\theta _t$ in the $(t\bar t)_{cm}$
frame.

The amplitudes for $q_i\overline{q_i}\rightarrow t\bar t$ in
the helicity
phase convention are easily obtained from those in Ref. [12].
The
associated
production density matrix elements are
\begin{equation}
\begin{array}{c}
\rho _{++}(q\bar q\rightarrow t\bar t)=\rho _{++,++}=\rho _{-
-,--} \\
=\frac{m_t^2}{9E_t^2}\sin ^2\theta _t
\end{array}
\end{equation}
\begin{equation}
\begin{array}{c}
\rho _{+-}(q\bar q\rightarrow t\bar t)=\rho _{+-,+-}=\rho _{-
+,-+} \\
=\frac 19(1+\cos ^2\theta _t)
\end{array}
\end{equation}
The normalization in these equations correspond to the hard
parton,
differential cross-sections
\begin{equation}
\frac{d\widehat{\sigma }}{dt}=\frac{\alpha _s^2}{s^2}(\rho
_{++,++}+\rho
_{--,--}+\rho _{+-,+-}+\rho _{-+,-+})
\end{equation}

\section{ADDITIONAL REMARKS}

The simpler stage-one spin-correlation
function$I(E_{W^{+}},E_{W^{-}})$ of
Ref. [5] directly follows from Eq.(69) by integrating
out $\widetilde{%
\theta _1}$ and $\widetilde{\theta _2}$%
\begin{equation}
I(E_{W^{+}},E_{W^{-}})
= \sum_{i} \{\rho _{+-}(q_i
\overline{q_i}\rightarrow t\bar t)^{prod}[\rho
_{++}\overline{\rho _{--}}%
+\rho _{--}\overline{\rho _{++}}] \\ +\rho
_{++}(gg\rightarrow t\bar
t)^{prod}[\rho _{++}\overline{\rho _{++}}+\rho _{--
}\overline{\rho _{--}}]\}
\end{equation}
where
\begin{equation}
\begin{array}{c}
\rho _{++}=1+\zeta
{\cal S}_W\cos \theta _1^t,\rho _{--}=1-\zeta {\cal S}_W\cos
\theta _1^t \\
\overline{\rho _{++}}=1-\zeta {\cal S}_W\cos \theta
_2^t,\overline{\rho _{--}%
}=1+\zeta {\cal S}_W\cos \theta _2^t
\end{array}
\end{equation}
However, using $I(E_{W^{+}},E_{W^{-}})$ the fractional
sensitivity for
measurement of $\zeta $ at the Tevatron at $2$ $TeV$ is only
$38\%$ versus $%
2.2\%$ by using $I(E_{W^{+}},E_{W^{-}},\widetilde{\theta
_1},\widetilde{%
\theta _2})$. The ``fractional sensitivity" is explicitly
defined by Eq.(36) in [6]. Similarly, at the LHC at $14$
$TeV$, the
fractional
sensitivity for measurement of $\zeta $ with $I_2$ is
$2.3\%$ versus $0.39\%
$ with $I_4$. This shows the importance of including the
analyzing power of
the second stage in the decay sequence, i.e. W decay-
polarimetry, c.f. Sec. 5.3.
It is also important to note that only the partial width and
the $\zeta$ helicity
parameter
appear in this stage-one spin-correlation function. To
measure the other helicity parameters $( \xi, \sigma, \ldots
)$, one needs to use stage-two W or b decay-polarimetry,
and/or other spin-correlation functions.

This use of W decay-polarimetry and $I_4$ to significantly
increase the analyzing powers does not directly make use of
the threshold-type kinematics at the Tevatron of the $q \bar
q \rightarrow t \bar t $ reaction.  See the series of papers
by Parke, Mahlon, and Shadmi [7] for spin-correlation
analyses which investigate threshold techniques.

Some modern Monte Carlo simulations do include
spin-correlation effects, for instance KORALB for $e^- e^+$
colliders[16].  The simple general structure and
statistical sensitivities of the S2SC function $I_4$ show
that spin-correlation effects should also be included in
Monte Carlo simulations for $p \bar p$, or $pp$, $\rightarrow
t \bar t X \rightarrow \ldots$.  In such a Monte Carlo it
should be simple and
straightforward to build in the
amplitudes for production of L-polarized and T-polarized
$W^{\pm}$'s
from distinct Lorentz-structure sources. Thereby,
spin-correlation techniques and the
results in
this paper can  be used for many systematic
checks.  For example, they could be used to experimentally
test the CP and T invariance ``purity" of detector components
and of the data analysis by distinguishing which coefficients
are or aren't equal between various experimental data sets
analyzed separately for the $t $ and $\bar t$ modes.

Assuming only $b_L$ couplings[17], a simple way for one to
use
a
Monte Carlo simulation to test for possible CP violation
is
to add an $S + P$ coupling (to the standard $V-A$ coupling)
in
the $t$ decay mode such that the $S+P$ contribution has an
overall complex coupling factor ``c" in the $t$ mode
and a complex factor ``d" in the $\bar t$ mode.  This will
generate a difference in modui and phases between the
$t, \bar t$ modes.  Then the 2 tests for CP violation
are
whether $|c| = |d|$, $arg(c) = arg(d)$ experimentally.

To be model independent and of greater use to
theorists,
experimental analyses should not assume a mixture
of only $V$ and $A$ current couplings in
top decays.  By
consideration of polarized-partial-widths there are several
fundamental quantities besides the chirality
parameter and the total partial width which can be directly
measured.  For example, there are three
logically independent tests for only $b_L$ couplings: $\xi
= 1$, $\zeta = \sigma$, and $\omega = \eta$ up to $O(m_b)$
corrections[18]. If $\tilde
T_{FS}$-violation were to occur, then the non-zero parameters
$\omega^{'} = \eta^{'}$
if there are only $b_L$ couplings.

\begin{center}
{\bf Acknowledgments}
\end{center}

We thank experimental and theoretical physicists for
discussions and assistance, in particular with respect to
matters specific to hadron colliders.  We thank Ming
Yang; and for computer services, John Hagan,
Christine Place-Sweet, and Mark Stephens. This work was
partially
supported by U.S. Dept. of Energy Contract No. DE-FG
02-96ER40291.

\newpage

\begin{center}
{\bf Table Captions}
\end{center}

Table 1: Analytic form of the helicity
parameters for $%
t\rightarrow W^{+}b$ decay for unique Lorentz couplings:  In
this
and following table, the mass ratios are denoted by $
w / t \equiv m_{w} / m_{t} $.  We do not
tabulate $\omega^{'}$ and $\eta^{'}$
because $\omega^{'}=\eta^{'}=0$ if either (i) there is a
unique
Lorentz coupling, (ii) there is no $\tilde T_{FS}$-violation,
and/or (iii) there is a  ``$V$ and $A, m_{b}=0$'' masking
mechanism,
see remark (5) in Sec. 1.

Table 2:  Analytic forms and numerical values of the
partial-width-intensities for polarized final states for
unique Lorentz couplings.

Table 3: Helicity parameters for $%
t\rightarrow W^{+}b$ decay to leading-order in the case of a
single
additional chiral coupling ($g_{\iota}$) which is small
relative to the standard $V-A$ coupling ($g_L$).  This table
is for the $V+A$ and for the $S \pm P$ couplings. The next
table is for additional tensorial couplings.   In this paper
$\cal RE$ ( $\cal IM$ ) denote respectively the real
(imaginary) parts of the quantity inside the parentheses.

Table 4: Same as previous table except this table is for
additional tensorial couplings.
Here $g_\pm = f_M \pm f_E$ involves $ k_t - p_b $
whereas
$\tilde{g}_{\pm}={g^+}_{T^+ \pm T_5^+}$ involves $ k_t +
p_b $, see Eqs.(22).  Here $m_t=$ mass of the $t$ quark.

\begin{center}
{\bf Figure Captions}
\end{center}

FIG. 1: The three angles $\theta _1^t$,$\theta _2^t$ and
$\phi $ describe
the first stage in the sequential-decays of the $(t\bar t)$
system in which $%
t\rightarrow W^{+}b$ and $\bar t\rightarrow W^{-}\bar b$.
From (a) a boost
along the negative $z_1^t$ axis transforms the kinematics
from the $t_1$
rest frame to the $(t\bar t)_{cm}$ frame and, if boosted
further, to the $%
\bar t_2$ rest frame shown in (b).

FIG. 2: The two pairs of spherical angles $\theta _1^t$,
$\phi _1^t$ and $%
\widetilde{\theta _a}$,$\widetilde{\phi _a}$ describe the
respective stages
in the sequential decay $t\rightarrow W^{+}b$ followed by
$W^{+}\rightarrow
j_{\bar d}j_u$, or $W^{+}\rightarrow l^{+}\nu $.  The
spherical angles $%
\widetilde{\theta _a}$, $\widetilde{\phi _a}$ specify the
$j_{\bar d}$ jet
(or the $l^{+}$) momentum in the $W^{+}$rest frame when the
boost is from
the $t_1$ rest frame. For the
hadronic $W^{+}$ decay mode, we use the notation that the
momentum of the charge $\frac{1}{3} e$ jet is denoted by
$j_{\bar d}$ and the momentum of the charge $\frac{2}{3} e$
jet by $j_{u}$.  In this figure, $\phi _1^t$ is
shown equal to
zero for simplicity of illustration.

FIG. 3: This figure is symmetric versus Fig. 2.  The
spherical angles $\widetilde{\theta _b}
$, $\widetilde{\phi _b}$ specify the $j_d$ jet (or the $l^{-
}$) momentum in
the $W^{-}$rest frame when the boost is from the $\bar t_2$
rest frame.

FIG. 4: The spherical angles $\widetilde{\theta
_1}$,$\widetilde{\phi _1}$
specify the $j_{\bar d}$ jet (or the $l^{+}$) momentum in the
$W^{+}$rest
frame when the boost is directly from the $(t\bar t)_{cm}$
frame. Similarly,
$\widetilde{\theta _2}$, $\widetilde{\phi _2}$ specify the
$j_d$ jet (or the
$l^{-}$) momentum in the $W^{-}$rest frame. The $W^{+}W^{-}$
production
half-plane specifies the positive $x_1$ and $x_2$ axes.

FIG. 5: Display of test for $\widetilde{T_{FS}}$ violation
using the
right-triangle relation, Eq.(60): First, side $a=\eta +\omega
$ is drawn with
its uncertainty $\delta _a$ and then the hypotenuse $c=\frac
12\sqrt{\left[
(1+\xi )^2-(\sigma +\zeta )^2\right] }$ is cast to form a
right-triangle. $c$%
's uncertainty is shown as $\delta _c$. A resulting non-zero
side $b=\eta
^{^{\prime }}+\omega ^{^{\prime }}$ would imply that
$\widetilde{T_{FS}}$ is
violated either dynamically or because of a fundamental
violation of
canonical $T$-invariance.

FIG. 6: Display of the $W^{+}$ energy-$W^{-}$ energy
correlation, $%
I_2^{cm}(\cos \theta _1^t,\cos \theta _2^t)$ as predicted by
the standard
model  for $pp\rightarrow t\bar tX$ (LHC). The contours shown
are for $10^6$ events over
$10$bins $ \cdot $ $10$bins (LHC).  This
saddle surface
peaks at $(\pm 1,\mp 1)$; and the levels range from $9,478$,
to $10,522$
with spacing $116$.  [At the Tevatron at $2 TeV$, the saddle
is inverted with  dips at $(\pm 1,\mp 1)$; with levels
ranging
from $294$ to $306$ with spacing $1.2$ for $3 \cdot 10^4$
events] .

FIG. 7: First of 8 figures showing the $\cos \theta _1^t,\cos
\widetilde{%
\theta _1}$ behaviour of the elements of the ``reduced''
composite-density-matrix $\rho
_{hh^{^{\prime }}}$. These also show the dependence as the
total
center-of-mass energy $E_{cm}$ is changed. This figure is for
$\rho _{++}$
and $E_{cm}=380$ $GeV$; the next figure is for $E_{cm}=450$
$GeV$. This
saddle surface peaks at about $(1,0)$, $(-1,-1)$; and the
levels range from $%
0.1300$ to $1.3923$ with spacing $0.1266$.

FIG. 8: The $\cos \theta _1^t,\cos \widetilde{\theta _1}$
behaviour of $%
\rho _{++}$  for $E_{cm}=450$ $GeV$. The surface peaks at
about $(1,0)$, and
falls towards the 3 corners; the levels range from  $0.1751$
to $1.3422$
with spacing $0.1220$.

FIG. 9: The $\cos \theta _1^t,\cos \widetilde{\theta _1}$
behaviour of $%
\rho _{--}$ for $E_{cm}=380$ $GeV$. The saddle surface peaks
at about $(-1,0)
$,$(1,-1)$; the levels range from $0.1231$ to $1.2274$ with
spacing $0.1227$.

FIG. 10: The $\cos \theta _1^t,\cos \widetilde{\theta _1}$
behaviour of $%
\rho _{--}$ for $E_{cm}=450$ $GeV$. The surface peaks at
about $(-1,1)$, $
(-0.5,-1)$; the levels range from $0.1404$ to $1.4002$ with
spacing $0.1400$.

FIG. 11: The $\cos \theta _1^t,\cos \widetilde{\theta _1}$
behaviour of $%
Re[\rho _{+-}]$ for $E_{cm}=380$ $GeV$. The surface peaks at
about $(-0.25,0.25)$
; the levels range from $-0.5392$ to $0.4179$ with spacing
$0.1063$.

FIG. 12: The $\cos \theta _1^t,\cos \widetilde{\theta _1}$
behaviour of $%
Re[\rho _{+-}]$ for $E_{cm}=450$ $GeV$. The surface peaks at
about $(-0.8,0.9)$;
the levels range from $-0.5960$ to $0.4490$ with spacing
$0.1161$.

FIG. 13: The $\cos \theta _1^t,\cos \widetilde{\theta _1}$
behaviour of $%
Imag[\rho _{+-}]$ for $E_{cm}=380$ $GeV$ for arbitrary
overall normalization $\omega^{\prime} = \eta^{\prime}$ $=1$.
The surface peaks
at about $%
(-0.5,0.5)$; the levels range from $-0.5025$ to $0.1293$ with
spacing $0.0702
$.

FIG. 14: The $\cos \theta _1^t,\cos \widetilde{\theta _1}$
behaviour of $%
Imag[\rho _{+-}]$ for $E_{cm}=450$ $GeV$ for arbitrary
overall normalization $\omega^{\prime} = \eta^{\prime}$ $=1$.
The surface peaks
at about $%
(-1,0.8)$; the levels range from $-0.7025$ to $0.2842$ with
spacing $0.1096$.


\begin{thebibliography}{333}
\bibitem{1} F. Abe, et. al. (CDF collaboration), Phys. Rev.
Lett. {\bf 74}, 2626(1995).
\bibitem{2} S. Abachi, et. al. (D0 collaboration), Phys. Rev.
Lett. {\bf 74}, 2632(1995).
\bibitem{3} P. Tipton, ICHEP'96-Warsaw Conference
Proceedings, p. 123, eds. Z. Ajduk and A.K. Wroblewski (World
Sci., Singapore, 1997).
\bibitem{JW} M. Jacob and G. Wick, Ann. Phys. (N.Y.) {\bf 7},
209(1959); K.-C. Chou, JETP {\bf 36}, 909(1959); M.I.
Shirokov, {\it ibid} {\bf 39}, 633(1960); J. Werle, Phys.
Lett. {\bf 4}, 127(1963); S.M. Berman and M. Jacob, SLAC
Report No.
43(1965); Phys. Rev. {\bf 139} B1023(1965); R.D. Auvil and
J.J. Brehm, {\it ibid.} {\bf 145}, 1152(1966). Very readable
treatments of the helicity formalism are in H. Pilkuhn, {\it
Interaction to Hadrons} (North-Holland, Amsterdam, 1967);
M.L. Perl, {\it High Energy Hadron Physics} (Wiley, N.Y.,
1974); A.D. Martin and T.D. Spearman, {\it Elementary
Particle Physics} (North-Holland, Amsterdam, 1970); M.L.
Goldberger and K.M. Watson, {\it Collision Theory} (Wiley,
N.Y. 1964); J. Werle, {\it Relativistic Theory of Reactions}
(North-Holland, Amsterdam, 1966); J.D. Richman, Caltech
Reports No. CALT-68-1148(unpublished), -1231(unpublished).
\bibitem{5} C.A. Nelson, Phys. Rev. {\bf D41}, 2805(1990); p.
259, in  M.
Greco(ed), {\it Results and Perspectives in Particle
Physics}, ( Editions Frontiers, France 1994).
\bibitem{PL} C.A. Nelson, B.T. Kress, M. Lopes, and T.P.
McCauley, SUNY BING 5/27/97,hep-ph/9706469.
\bibitem{7} V. Barger, J. Ohnemus and R.J.N. Phillips, Intl.
J. Mod. Phys. {\bf A4}, 617(1989); G.L. Kane, G.A. Landinsky,
and C.P. Yuan, Phys.
Rev. {\bf D45}, 124(1991); R.H. Dalitz and G.R. Goldstein,
{\it ibid.}{\bf D45},  1531(1992); M. Jezabek and J.H. Kuhn,
Phys. Lett. {\bf B329}, 317(1994); G.A. Landinsky and C.P.
Yuan, Phys.
Rev. {\bf D49},  4415(1994); E. Malkawi and C.P.
Yuan,{\it ibid.}{\bf D50},  4462(1994); G. Mahlon and S.
Parke, {\it ibid.}{\bf D53}, 4886(1996); hep-ph/9706304; S.
Parke and Y. Shadmi, Phys. Lett. {\bf B387},  199(1996); T.
Stelzer and S. Willenbrock,{\it ibid.}{\bf B374}, 169(1996);
A. Brandenburg,{\it ibid.}{\bf B388}, 626(1996);
B. Grzadkowski and Z. Hioki,{\it ibid}{\bf B391}172(1997),
hep-ph/9610306; D. Chang, S.-C. Lee and A. Sumarokov, Phys.
Rev. Lett.{\bf 77},  1218(1996); A.P. Heinson, A.S. Belyaev,
and E.E. Boos, hep-ph/9612323; K. Cheung, Phys. Rev.{\bf
D55},  4430(1997).
\bibitem{8} J.F. Donoghue and G. Valencia, Phys. Rev. Lett.
{\bf 58},  451(1987); C.A. Nelson, Phys. Rev. {\bf D30}, 1937
(1984); J.R. Dell'Aquila and C.A. Nelson,{\it ibid.}{\bf
D33},  80,  101(1986); W. Bernreuther, J.P.
Ma, T. Schroder, and Pham, Phys. Lett. {\bf B279},
389(1992); W. Bernreuther, O. Nachtmann, P. Overmann, and T.
Schroder,
Nucl. Phys. {\bf B388},  53(1992); C.R. Schmidt and M.E.
Peskin, Phys. Rev. Lett. {\bf 69},  410(1992); D. Chang and
W.-Y. Keung, Phys. Lett.{\bf B305}, 261(1993).
\bibitem{9} J.P. Ma. and A. Brandenburg, Z. Phys. {\bf
C56},  97(1992); T. Arnes and L.M. Sehgal, {\it ibid.} {\bf
302},  501(1993); D. Atwood, G. Eilam, A. Soni, R. Mendel and
R. Migneron, Phys. Rev. Lett. {\bf 70},  1364(1993); R. Crux,
B. Gradkowski and J.F. Gunion, Phys. Lett.{\bf B289},
440(1992); B. Gradkowski, B. Lampe, and K.J. Abraham, hep-
ph/9706489.
\bibitem{10} C.A. Nelson, Phys. Rev. {\bf D53},  5001(1996);
Phys. Lett. {\bf B355},  561(1995).
\bibitem{11}C.A. Nelson, H.S. Friedman,
S. Goozovat, J.A. Klein, L.R. Kneller, W.J. Perry, and S.A.
Ustin, Phys. Rev. {\bf D50},  4544(1994).
\bibitem{12} C.A. Nelson, Phys. Rev. {\bf D43}, 1465(1991).
\bibitem{13} C.A. Nelson, Phys. Rev. {\bf D30}, 1937(1984);
J.R. Dell'Aquila and C.A. Nelson, {\it ibid.}{\bf 33},
101(1986).
\bibitem{14} F. Abe, et. al. (CDF collaboration), Fermilab-
PUB-97/075-E.
\bibitem{15} M. Gluck, J.F. Owens, and E. Reya, Phys. Rev.
{\bf D17}, 2324(1978); H. Georgi, S.L. Glashow, M.E.
Machacek, and D.V. Nanopoulos, Annals of Phys. {\bf 114},
273(1978).
\bibitem{16} S. Jadach and Z. Was, Comp. Phys. Commun. {\bf
36}, 191(1985); {\it ibid.}{\bf 64} 275(1991); {\it
ibid.}{\bf 85} 453(1995).
\bibitem{17} Historically, for a complete Lorentz-invariant
characterization of the
charged current, there have been two popular choices for the
minimal sets of
couplings: Express the vector matrix element $<b|v^\mu (0)|t
>$ either in
terms of $V,f_M,$and $S^{-}$(recall  $S^{-}$ doesn't
contribute
to the on-shell $W^+$ mode) or of $V,S$ and $S^{-}$.
Correspondingly,
express the
axial $<b|a^\mu (0)|t>$ either in terms of $A,f_E,$and $P^{-
}$(  $P^{-}$%
doesn't contribute to the on-shell $W^+$ mode) or of $A,P$
and $P^{-}$.
So if there are only $b_L$ couplings, then the chiral
combinations of $V-A
$ and $S+P$ can contribute significantly to the  $W^+$ mode
since $m_b/m_t, m_b/m_w \cong 0.$ To include $b_R$ couplings,
one would add $%
V+A$ and $S-P$ via an additional ``$g_R\gamma ^\mu (1+\gamma
_5)+e(k+p)^\mu
(1-\gamma _5)$'' for  $t$  and
``$\bar{g_R} \gamma
^\mu (1+\gamma _5)+f(k+p)^\mu (1-\gamma _5)$''
for $\bar t$  where `` $e$ '' and `` $f$ '' are different
complex
coupling
factors. In each case this gives the expected number of
independent
variables. Measurement of the overall relative phase of the
$\lambda_b= - \frac{1}{2}$ and
$\lambda_b= \frac{1}{2}$ couplings ( and for the $\lambda_b=
\pm \frac{1}{2}$ couplings of the anti-$b$ quark) by S2SC's
using $b$ quark-polarimetry is considered in Ref.[6].
\bibitem{18} The corrections for $m_b = 4.5GeV$ are given
numerically in Table 3 of Ref. [6] and follow analytically
from Eqs.(25-28) in the present paper.
\end{thebibliography}
\end{document}